\documentclass{aa}  

\usepackage{graphicx}
\usepackage{txfonts}
\usepackage{amsmath}
\usepackage{xcolor}
\usepackage{adjustbox}
\begin{document}

\title{Spectroscopic characterization of the protocluster of galaxies around 7C~1756+6520 at $z \sim 1.4$
\thanks{Based on data acquired using the Large Binocular Telescope (LBT). 
The LBT is an international collaboration among institutions in the United States, Italy and Germany. 
LBT Corporation partners are: The University of Arizona on behalf of the Arizona university system; 
Istituto Nazionale di Astrofisica, Italy; LBT Beteiligungsgesellschaft, Germany, representing the
Max-Planck Society, the Astrophysical Institute Potsdam, and Heidelberg University; 
The Ohio State University, and The Research Corporation, on behalf of The University of Notre Dame, 
University of Minnesota and University of Virginia.
}
}


\author{V.~Casasola\inst{1},
L.~Magrini\inst{1},
F.~Combes\inst{2},
E.~Sani\inst{3},
J.~Fritz\inst{4},
G.~Rodighiero\inst{5},
B.~Poggianti\inst{6},
S.~Bianchi\inst{1},
E.~Liuzzo\inst{7}
}

\institute{INAF -- Osservatorio Astrofisico di Arcetri, Largo E. Fermi 5, 50125, Firenze, Italy\\
\email{casasola@arcetri.astro.it}
\and
LERMA, Observatoire de Paris, CNRS, PSL Univ., Sorbonne Univ. and College de France,  Paris, France
\and
European Southern Observatory, Alonso de Cordova 3107, Casilla 19, Santiago 19001, Chile
\and
Instituto de Radioastronom\'\i a y Astrof\'\i sica, UNAM, Campus Morelia, A.P. 3-72, C.P. 58089, Mexico
\and
Dipartimento di Fisica e Astronomia, Universit{\`a} di Padova, vicolo dell'Osservatorio 3, 35122, Padova, Italy
\and
INAF-Astronomical Observatory of Padova, vicolo dell'Osservatorio 5, I-35122 Padova, Italy
\and
INAF -- Istituto di Radioastronomia, Via Piero Gobetti 101, 40129 Bologna, Italy
}

\date{Received; accepted}

\titlerunning{Galaxy cluster 7C~1756+6520 at $z \sim 1.4$}
\authorrunning{V.~Casasola et al.}

 
\abstract
{}
{
The aim of this paper is the spectroscopic study of 13 galaxies belonging to the field of the protocluster 
associated with the radio galaxy 7C~1756+6520 at $z = 1.4156$.
In particular, we focus on the characterization of the nuclear activity.         
}
{This analysis has been performed on rest-frame optical spectra taken with the Large Binocular Telescope, 
using the spectrograph LUCI, which is
operative in the near-infrared domain.
The adopted spectral coverage
allowed us to observe emission lines such as H$\alpha$, H$\beta$, [O {\sc iii}]$\lambda$~5007~\AA, and [N {\sc ii}]$\lambda$~6583~\AA\
at the redshift of the central radio galaxy.
We observed the central part of the protocluster, which is suitable to include the radio galaxy, several spectroscopically confirmed 
active galactic nuclei (AGN) belonging to the protocluster, and other objects that might be members of the protocluster.
}
{
For four previously identified protocluster members,
we derived the redshift by detecting emission lines that have never detected before for these galaxies.
We identified a new protocluster member and eight new possible protocluster members. 
The stacked spectrum of the galaxies in which we detected the [O {\sc iii}]$\lambda$~5007~\AA\ emission line
revealed the second line of the [O {\sc iii}] doublet at 4959~\AA\ and the H$\beta$ line, which confirms that they belong to the protocluster. 
By collecting all members identified so far in this work and other members from the literature, 
we defined 31 galaxies, including the central radio galaxy, around the redshift $1.4152 \pm 0.056$. This corresponds 
to peculiar velocities $\lesssim$5000~km~s$^{-1}$ with respect to the radio galaxy.
The position versus velocity phase-space diagram suggests that three AGN
of the protocluster and the central radio galaxy might be a virialized population that has been coexisting for a long time 
in the densest core region of this forming structure.
This protocluster is characterized by a high fraction of AGN ($\sim$23$\%$). 
For one of them, AGN1317, 
we produced two so-called Baldwin, Phillips \& Terlevich (BPT) diagrams.  
The high fraction of AGN and their distribution within the protocluster seem to be consistent with predictions 
of some theoretical models on AGN growth and feedback. These models are based on galaxy interactions and ram pressure 
as triggers of AGN activity.
} 
{
The high fraction of AGN belonging to the protocluster suggests that they were likely triggered at the same time, maybe by the ongoing formation of the protocluster.
Observations of AGN in this protocluster and in other distant clusters will help clarifying
whether the resulting high fraction of AGN is unusual or typical for such structures at high redshift.
Our next step will be analyses of previously acquired high-resolution radio data of the central radio galaxy to derive information on the nature of the radio galaxy and connect it with its cosmic evolution.
}


\maketitle
%

\section{Introduction}
\label{sec:intro}
Galaxy clusters provide an efficient tool for deriving cosmological parameters and for studying galaxy formation and evolution.   
As the largest collapsed structures in the Universe with total masses of up to 10$^{15}$~M$_{\odot}$ \citep[e.g.,][]{arnaud09}, the cosmic history of
galaxy clusters is sensitive to key cosmological parameters \citep[e.g.,][]{voit05,vikhlinin09,stern10}.
Galaxy clusters allow us to study large populations of early-type galaxies, which provide stringent tests on galaxy evolution
models in the current hierarchical formation paradigm \citep[][]{renzini06}, because they are the most massive galaxies with the
oldest stellar populations (at least out to redshift $z \sim 2$). 

Over the past three decades, considerable effort has been devoted to discovering ever more distant galaxy clusters using different
observational methods \citep[see][for a review]{rosati02}.
X-ray selection of galaxy clusters has been central in several studies, as it naturally provides gravitationally bound systems  with a relatively simple selection function.
However, only few clusters are currently confirmed at $z > 1.$ 
Examples of distant, confirmed galaxy clusters identified through the extended X-ray emission of the intracluster medium are 
XMMU~J2235.3-2557 at $z = 1.39$ \citep[][]{mullis05,lidman08,rosati09}, XMMXCS~J2215.9-1738 at $z = 1.46$ \citep[][]{stanford06,hilton07},  
2XMM~J083026.2+524133 at $z = 0.99 $ \citep[][]{lamer08},
XDCP~J0044.0-2033 at $z = 1.58$ \citep[][]{fassbender14}, and
XMMUJ2235-2557 at $z = 1.39$ \citep[][]{chan16}.
However, the X-ray identification of candidate clusters is very difficult at $z > 1$ since the surface brightness of the extended X-ray emission fades as $(1 + z)^{4}$.

Another method for finding galaxy clusters relies upon the observation that all rich clusters, at all redshifts observed so far, appear to have a red sequence of early-type galaxies 
\citep[][]{gladders00}.
The colors of such galaxies are quite distinct as a result of the strong 4000~\AA\ break 
in their spectra.
However, this break shifts into the near-infrared (NIR) at $z > 1.5$ and the colors can become degenerate.
The fundamental advantage of this approach, the so-called red sequence method,
 is that with appropriate filters, cluster elliptical galaxies at a given redshift are redder than all normal galaxies at lower redshifts. 
We stress that the red sequence method could be biased at high redshift since galaxies tend to become bluer and younger at high-$z$.  
Despite its biases,
this technique has given notable results using \textit{Spitzer Space Telescope} IRAC bands, and several galaxy clusters found in this way have been spectroscopically confirmed at $z > 1.3$
\citep[e.g.,][]{eisenhardt08,papovich08,wilson09}.
More recently, similar results have been obtained using the IR photometry in the UKIDSS Ultra-Deep Survey
field \citep[][]{lawrence07} and $Hubble$ spectroscopy \citep[e.g.,][]{Lee-Brown17}.   

During the past two decades, a new technique for detecting galaxy clusters at $z > 1$ has been to study the immediate surroundings of high-redshift radio galaxies 
(hereafter H$z$RGs). 
It is now well established that the host galaxies of powerful radio sources are among the most massive galaxies in the Universe 
\citep[M$_{\rm star} > 10^{11}$~M$_{\odot}$,][]{seymour07}.
Because of their high masses, radio galaxies indeed represent excellent signposts to pinpoint the densest regions of the Universe out to very high redshifts \citep[e.g.,][]{stern03}.
Some examples are the overdensities of Ly$\alpha$ and H$\alpha$ emitters around H$z$RGs at $2.1 < z \leq 5.2$ 
\citep[][]{kurk04a,miley04,venemans05,venemans07}.
These overdensities are likely to be progenitors of present-day (massive) galaxy clusters. 
However, the Ly$\alpha$ and H$\alpha$ emitters found in these dense environments are small, faint, blue objects, likely young star-forming galaxies
with masses of a few $\times$10$^{8}$~M$_{\odot}$ \citep[][]{overzier08}, which may represent only a small fraction of the number of cluster galaxies 
and the total mass of the cluster.

Interestingly, overdensities at the highest redshifts often have a filamentary nature and extend beyond 
the projected dimension of $\sim$2 Mpc \citep[e.g.,][]{pentericci98,carilli02,croft05}.
\citet{carilli02} have studied the filaments in the field of the H$z$RG PKS~1138-262 at $z = 2.1$ in detail,
but did not detect any extended X-ray emission, indicating that this structure has not yet had sufficient time to virialize.
However, \citet{kurk04b} has shown that some segregation has occurred in this overdensity: the H$\alpha$ emitters,
tracing the more evolved population, are more centrally concentrated than the younger Ly$\alpha$ emitters.    
This suggests that the missing link between these protoclusters and the classical X-ray confirmed
clusters found out to $z \sim 1.4$ \citep[e.g.,][]{mullis05,stanford06} apparently occurs in the redshift range 
$1.4 < z \leq 2$.
This redshift range is therefore particularly interesting for identifying clusters at a redshift beyond which the classical selection
techniques are sensitive, but at a redshift where clusters are already partly virialized and a core of older, massive galaxies is in place.

Although the method of studying the immediate surroundings of H$z$RGs has successfully found overdensities of red galaxies at $z > 2$, it has been challenging 
to spectroscopically confirm their association with the H$z$RGs \citep[e.g.,][]{venemans05,kodama07,doherty10}.
A few studies have also applied related methods to H$z$RGs with  slightly lower redshift \citep[e.g.,][]{best03,stern03,galametz09a,galametz10,franck15,cooke16}
and found overdensities of extremely red galaxies in the environments of H$z$RGs at $z \sim 1.5$.
At such high redshifts, these overdensities are suspected to still be forming and to be not yet (completely) bound, and the term ``protoclusters'' is commonly used 
to describe such systems.
This last selection technique, based on the surroundings of H$z$RGs, has permitted the discovery of the galaxy protocluster 
on which this paper is focused.
 
An important aspect in the study of clusters is the demographics and distribution of AGN within clusters and their evolution
with redshift.
The AGN population of a cluster has indeed important implications for the AGN fueling processes and how tightly black holes 
(BHs) at the centers of cluster galaxies and galaxies coevolve.    
Numerous lines of evidence suggest that there is coevolution between the growth of the BHs and the formation of stars in galaxies.
Perhaps the most striking result is the similar rate of evolution of the emissivity from AGN and star
formation from $z \sim 2$ to the present epoch \citep[e.g.,][]{boyle98,franceschini99,merloni04,silverman08}.

AGN in clusters relative to the field provide some valuable additional observational constraints 
to the physical processes that affect the availability and transport of the 
cold gas that serves as the primary fuel for the central BH (and SF).
These processes include the removal of cold gas through ram-pressure stripping \citep{cowie77}, 
evaporation by the hot interstellar medium \citep[ISM][]{gunn72}, tidal effects due to the cluster
potential \citep{farouki81,merritt83} and other galaxies \citep{richstone76,moore96}, 
and gas starvation due to the absence of new infall of cold gas \citep{larson80}.
These physical processes have been invoked to explain the relative absence of luminous, 
star-forming galaxies, the scarcity of substantial reservoirs of cold gas, and the large
fraction of relatively quiescent, early-type galaxies in local clusters 
\citep[e.g.,][]{gisler78,dressler80,giovanelli85,dressler99}.

Moreover, luminous AGN are rarer in local cluster galaxies than in field galaxies, 
although these results are based on works that do not sample the densest regions of clusters well
\citep[e.g.,][]{kauffmann04,popesso06}, while less luminous AGN appear to be present 
in comparable numbers 
\citep[e.g.,][]{martini06,haggard10,marziani17}. 
The most frequently invoked process for the fueling of the most luminous AGN is the merger of two gas-rich galaxies 
\citep[e.g.,][]{barnes92}, and the relative lack of both cold gas and major mergers is a reasonable 
explanation for the nearly complete absence of quasars hosted by cluster galaxies.  
For less luminous AGN, the case is less clear because an increasing number of physical processes
such as minor mergers, galaxy harassment, various types of bars, and stellar mass loss 
could also play a role \citep[e.g.,][]{vivi08,vivi11,francoise13,santi14}.

In addition to a local comparison between AGN in different environments, measurement of the evolution of the AGN population
in clusters can constrain the formation epoch for their supermassive BHs and the extent of their coevolution with the cluster 
galaxy population. 
The first quantitative evidence for a substantial increase in the cluster AGN fraction with redshift was presented by \citet{eastman07},
who compared the fraction of spectroscopically confirmed AGN of similar X-ray luminosities in low- and high-redshift clusters 
\citep[see also][]{galametz09b}. 
\citet{martini09} found that the AGN fraction in cluster increases as $(1 + z)^{5.3}$ 
\citep[see also][]{fassbender12,rumbaugh12,tanaka13}, and it is quite similar to the  
evolution of the fraction of star-forming galaxies in clusters, $(1 + z)^{5.7}$ \citep[e.g.,][]{haines09}. This
suggests that AGN and star-forming galaxy populations evolve at similar
rates in clusters.

The evolution of the AGN fraction in galaxy clusters appears to be substantially
greater than the evolution of the AGN fraction in the field, where 
the AGN fraction increases by only about a factor of two from $z \sim 0.5$ to $z \sim 1.2$ 
\citep[e.g.,][]{alonso-herrero08,bundy08}, 
which is several times smaller than the increase of AGN in clusters.
This relative evolution appears to be broadly consistent with the behavior of star-forming
galaxies in the same redshift range.
\citet{elbaz07} showed that the fraction of galaxies that are star-forming
is correlated with local galaxy density at $z \sim 1$, which is a
reversal of the anticorrelation observed in the Local Universe.
Nevertheless, a direct comparison between field and cluster surveys is complicated because 
the relevant studies often employ different selection criteria, such as luminosity in some bands, an
estimate of the stellar mass, and different AGN luminosity limits to establish their host galaxy and AGN samples. 
These selection criteria are important because the AGN fraction above a given luminosity limit depends on stellar mass
\citep[e.g.,][]{heckman04,sivakoff08,aird12}, and the evolution of the X-ray luminosity function indicates that more luminous AGN 
were proportionally more abundant at higher redshift \citep[the so-called AGN downsizing phenomenon,][]{barger05,hasinger05}.

In this work, we present the spectroscopic study of 13 galaxies identified in the field of the protocluster that is
associated with the radio galaxy 7C~1756+6520 at $z = 1.4156$ \citep[][]{galametz09a,galametz10}.      
In particular, we focus on the AGN population belonging to this protocluster.
The galaxies we observed comprise both known protocluster members and 
possible protocluster members.
This analysis has been performed on rest-frame optical spectra carried out with the
Large Binocular Telescope (LBT), and was stimulated by previous results we obtained for this protocluster 
with LBT \citep[][hereafter M12]{laura12a} and the IRAM Plateau de Bure Interferometer \citep[PdBI,][]{vivi13}.

The  paper is structured as follows. 
In Sect.~\ref{sec:cluster} we present the main properties of protocluster 7C~1756+6520, and 
in Sects.~\ref{sec:obs} and \ref{sec:reduction} we describe LBT observations and the data reduction, respectively.
The results on the galaxy population and AGN population belonging to protocluster 7C~1756+6520 
are collected and discussed in Sect.~\ref{sec:results} and \ref{sec:focus-agn}, respectively, 
and the comparison between observations and predictions of some theoretical models is presented in
Sect.~\ref{sec:theory}.
Finally, we summarize our work in Sect.~\ref{sec:conclusions}.
We assume a $\Lambda$CDM cosmology with $H_{0} = 70$~km~s$^{-1}$~Mpc$^{-1}$, $h = H_{0}/100$,
$\Omega_{\rm 0} = 0.3$, and
$\Omega_{\Lambda} = 0.7$.
With these values, 1\arcsec\ corresponds to $\sim$8.4 kpc at $z \sim 1.4$.

\section{The target: a spectroscopically confirmed protocluster at $z \sim 1.4$}
\label{sec:cluster}
\citet{galametz09a,galametz10} have isolated and spectroscopically confirmed with the optical Keck/DEep Imaging 
Multi-Object Spectrograph (DEIMOS) the overdensity of galaxies associated with the radio galaxy 7C~1756+6520 at 
$z = 1.4156$. 
The radio galaxy was initially reported to be at $z = 1.48$ by \citet{lacy99} based
on the identification of a single, uncertain emission feature. 
In addition to the central radio galaxy, 21 galaxies have been confirmed with spectroscopic redshifts 
consistent with that of 7C~1756+6520 \citep[][hereafter G10]{galametz10}. 
The spectroscopic identifications of G10 were based mainly on the detection of the [O {\sc ii}]$\lambda$~3727~\AA\ emission line,
and more rarely on that of the [Ne {\sc v}]$\lambda$~3426~\AA\ and [Ne {\sc iii}]$\lambda$~3869~\AA\ emission lines. 
In the field around the radio galaxy, the velocity dispersion is rather large, up to $\sim$13000~km~s$^{-1}$, 
and the ensemble has not yet relaxed into one large structure. 
For these reasons, the overall structure would be better defined as a galaxy overdensity.
However, the group of galaxies associated with 7C~1756+6520 is called indiscriminately a galaxy cluster, 
a protocluster of galaxies, or a galaxy overdensity.     
Throughout the paper
we adopt the term overdensity or protocluster to identify the galaxy structure around 7C~1756+6520.

Two distinct smaller subgroups have been identified within this galaxy overdensity: 
one of eight galaxies (including the radio galaxy) centered around 7C~1756+6520 at $z \sim 1.42$, 
{and a more compact group composed of four galaxies at $z \sim 1.44$.}
Galaxies belonging to both subgroups are within 2~Mpc from the radio galaxy, 
while most other galaxies belonging to the large-scale overdensity 
lie more than~2 Mpc away from the radio galaxy (see Fig.~\ref{fig:cluster} below for the spatial distribution of sources in the field of the protocluster). 
Seven of the spectroscopically confirmed galaxies of the protocluster, including the central radio galaxy, have been classified as AGN candidates 
by \citet{galametz09a} using the {\it Spitzer}/IRAC color-color selection of \citet{stern05} \citep[see also][]{lacy04}.
The nature of these AGN candidates
has subsequently beenconfirmed by G10, who found AGN signatures in their spectra, such as strong, 
broad Mg~{\sc ii}$\lambda$~2800~\AA\ emission lines.
Three of these AGN are located within 1\farcs5 of the radio galaxy, which adds evidence that a significant number of members of galaxy clusters at $z > 1$ 
are AGN, and that they lie preferentially near the cluster center \citep[][G10]{galametz09b}.   

Our group has previously studied a sample of star-forming galaxies associated with protocluster 7C~1756+6520 
with observations at the LBT 
using the NIR 
spectrograph LUCI (M12).
The aim of our previous work was to derive the star formation rate, metallicity, and stellar mass of these galaxies, and 
locate them in the plane of the so-called fundamental metallicity relation (FMR) by 
combining our spectroscopic observations and the literature photometric data. 
The FMR is known not to evolve with redshift up to $z = 2.5$ for field galaxies, but it is still poorly explored 
in rich environments at low and high redshifts \citep[e.g.,][]{mannucci10,leslie12,laura12b,hunt16}. 
We found that the properties of the star-forming galaxies in protocluster 7C~1756+6520 
are compatible with the FMR, suggesting that the effect of the environment on galaxy metallicity 
at this early epoch of cluster formation is marginal. 
We also reported the spectroscopic analysis of the most luminous AGN in the protocluster,  
AGN1317, which located in the neighborhood of the central radio galaxy at a projected distance 
of $\sim$780 kpc. We detected a strong gas outflow 
in the [O~{\sc iii}]$\lambda\lambda$~4959,
5007~\AA\ doublet that reaches velocities of $>$1000~km~s$^{-1}$ and might be driven by the AGN radiation pressure. 
In addition, M12 have spectroscopically identified a new protocluster member, called MSC2, at $z \sim 1.45$ (see Table~1 in M12). 

We also studied the molecular gas content of AGN1317.
Although the number of CO detections of galaxies that lie in overdensities at $z > 1$ and may be signs of the build-up of cluster
galaxy populations, that is, protoclusters, was (at the epoch of our CO observations in 2013) and still is small 
\citep[e.g.,][]{casey16,wang16,noble17,rudnick17,castignani18,hayashi18,strazzullo18}, 
almost all of the CO detections of protocluster members are physically associated
with H$z$RGs 
\citep[e.g.,][]{emonts14,dannerbauer17}.
Motivated by these findings, we searched for molecular gas content in AGN1317.
With the IRAM PdBI, we indeed detected CO emission in AGN1317 and 
derived a substantial molecular gas mass of $1.1 \times10^{10}$~M$_{\odot}$ \citep{vivi13}, 
which is comparable to that found in massive submillimeter galaxies at $z \sim 1 - 3.5$ \citep[e.g.,][]{greve05}
and star-forming galaxies at $z \sim 1.5 - 3$ \citep[e.g.,][]{carilli13}.

\begin{figure*}
\centering
\includegraphics[width=0.5\textwidth, angle=-90]{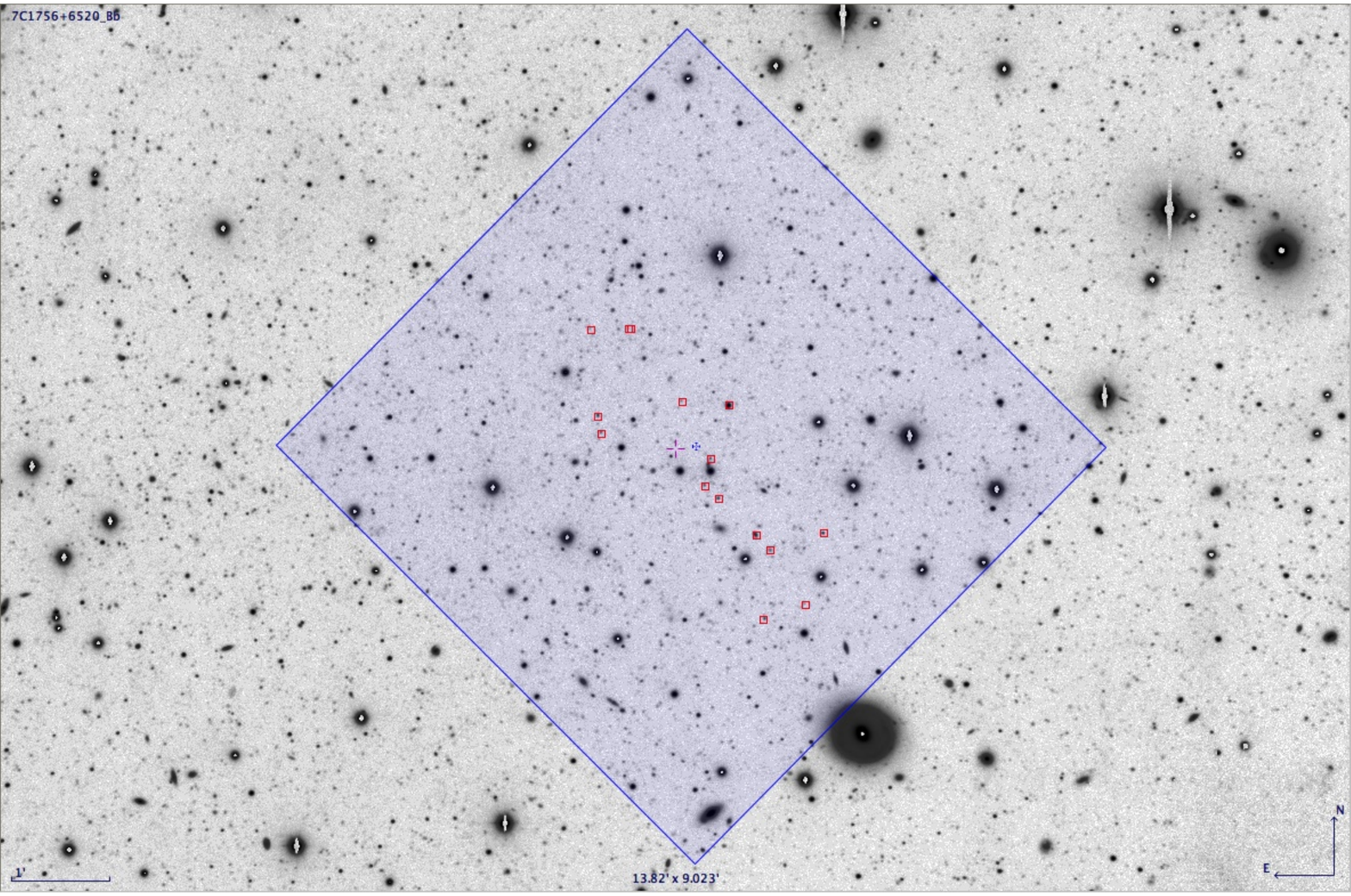}
\caption{
Location of the mask adopted for our observations at LBT plotted on the $B$-band image (NOAO).
The mask (in cyan) was centered at $\rm RA_{J2000}$~=~17$^{\rm h}$~56$^{\rm m}$~59$^{\rm s}$, 
$\rm Dec_{J2000}$~=~65$^{\circ}$~18$^{\prime}$~42$^{\prime\prime}$ with a rotation angle with respect to north 
of 45$^{\circ}$.
This mask allowed us to observe the central part of the protocluster, including the radio galaxy.}
\label{fig:location}
\end{figure*}

\begin{figure*}
\centering
\includegraphics[width=0.99\textwidth]{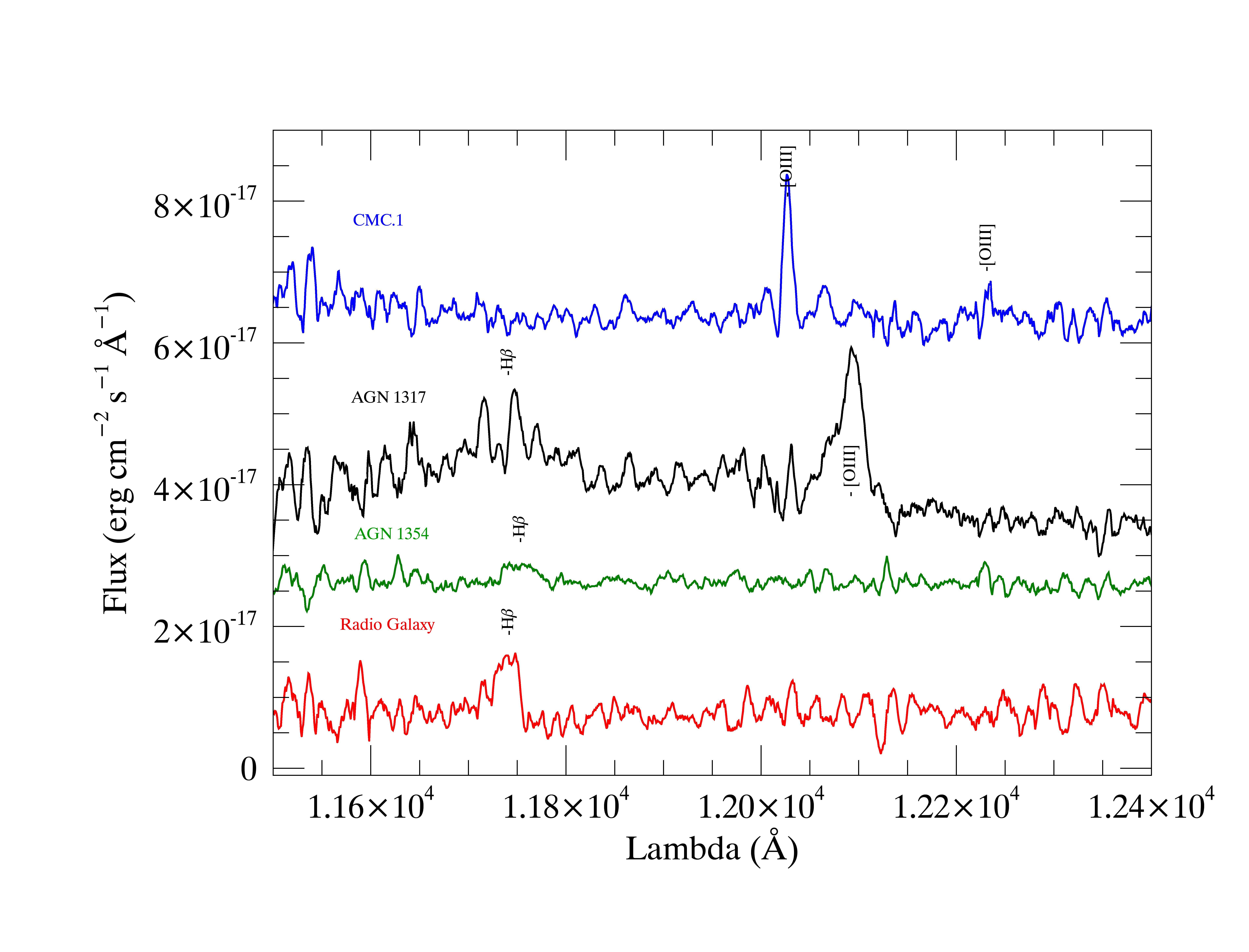}
\vspace{-1.5cm}
\caption{
1D spectra obtained 
in the $J$ wavelength range
for the following confirmed proto-cluster members: the radio galaxy, AGN1354, AGN1317, CMC~1
(see the first portion of Table~\ref{tab:z}, `Protocluster members'). 
The most prominent emission lines, H$\beta$ and  [O~{\sc iii}]$\lambda\lambda$~4959, 5007~\AA,  are labeled.  
The protocluster member AGN.1110 is not plotted (see the text for details).   
}
\label{fig:1DJ}
\end{figure*}

\begin{figure*}
\centering
\includegraphics[width=0.99\textwidth]{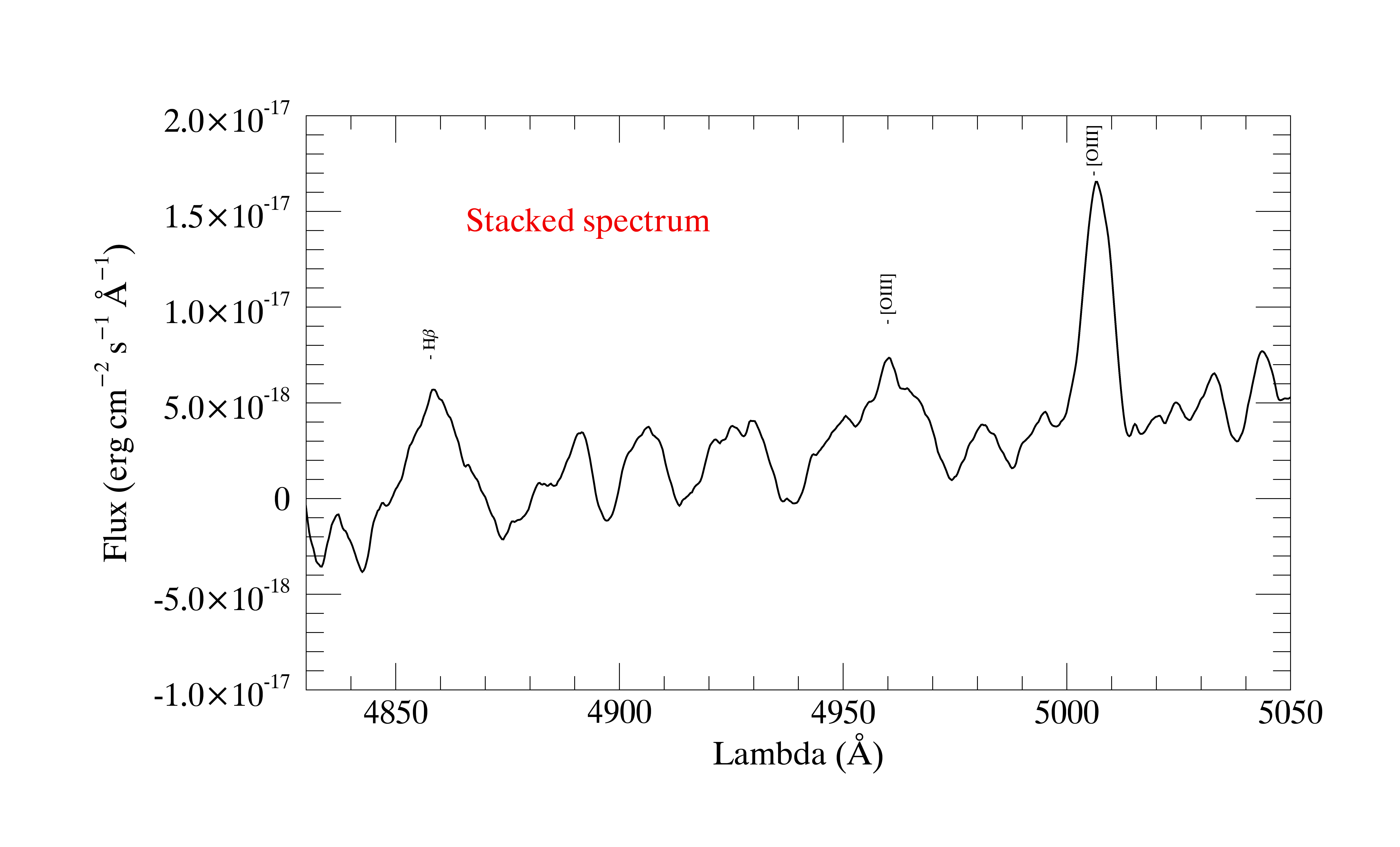}
\vspace{-1.0cm}
\caption{
1D stacked rest-frame spectrum in the $J$ wavelength range of the possible protocluster members where we detected the [O~{\sc iii}]$\lambda$~5007~\AA\ line
(CMC 2 - CMC 9, except for CMC 7, where we identified the detected emission line as the H$\beta$, see Table~\ref{tab:z}).  
}
\label{fig:1DJ-stack}
\end{figure*}

\section{Observations}
\label{sec:obs}
The observations presented here 
were taken in 2014 with the LBT, located on Mount Graham, Arizona 
\citep{hill06}, 
using the instrument LUCI 1 of the NIR spectrograph LUCI 
\citep[][]{ageorges10,seifert10}.
LUCI 1 is mounted on the bent Gregorian focus of the left primary mirror and has a wavelength coverage from 8500 to 25~000~\AA.
LUCI is available in imaging, long-slit, and 
multi-object spectroscopy (MOS) modes\footnote{
The other NIR instrument of LUCI is LUCI 2. This was not used in our observations.
}.

Taking advantage of the field of view (FoV) of LUCI 1 (the useful FoV = $4^{\prime} \times 2\farcm8$\footnote{The full FoV = $4^{\prime} \times 4^{\prime}$.}) 
in MOS mode, 
we obtained simultaneous deep spectroscopy of the AGN
belonging to the galaxy overdensity associated with radio galaxy 7C~1756+6520.
Figure~\ref{fig:location} shows the location of the mask we adopted for observations.
The mask (in cyan) was centered at $\rm RA_{J2000}$~=~17$^{\rm h}$~56$^{\rm m}$~59$^{\rm s}$, 
$\rm Dec_{J2000}$~=~65$^{\circ}$~18$^{\prime}$~42$^{\prime\prime}$ with a rotation angle with respect to north 
of 45$^{\circ}$, and this allowed us to observe the central part of the protocluster, including the radio galaxy.
The main targets of our observations were the AGN, spectroscopically confirmed by G10. 
Our list of highest priority targets includes AGN1317, the brightest galaxy, the central radio galaxy 7C~1756+6520, 
and two other AGN, AGN1110 and AGN1354. 
We also included several other objects in the mask design that might be protocluster members based on their location, magnitude, 
and angular size. 

To detect the most important emission lines in the rest-frame optical spectrum of AGN and of emission-line galaxies, 
we selected the grating G210-HiRes with the $J$ and $H$ wavelength ranges. 
The resulting spectral coverage, optimized to center the brightest emission lines of the spectra,  is
$11200-12600~\AA$ and 
$14800-16700~\AA$ in the $J$ and $H$ wavelength range, respectively. 
This allowed us to include several emission lines at the nominal redshift of the central H$z$RG of $z \sim 1.41$: 
H$\beta$ was observed at 
$11740~\AA$, 
[O {\sc iii}]$\lambda$~5007~\AA\ at 12080~\AA,
H$\alpha$ at 15820~\AA,
and [N {\sc ii}]$\lambda$~6583~\AA\ at 15870~\AA. 
The spectroscopic resolving power of G210-HiRes is R~=~8460 and R~=~7838 in the $J$ and $H$ wavelength range, respectively,  
meaning that with a slit width of 1$^{\prime\prime}$, the spectral resolution is 
$\sim$2.8~\AA\ and 
$\sim$4~\AA, respectively, which is suitable for separating  emission lines and sky lines. 

To perform the sky subtraction, we adopted the \textit{nodding} procedure, in which the object of interest is observed at different positions along the slit. 
The exposure was 5~hr in the $J$ wavelength range, and for technical reasons, it was only 150~s in the $H$ wavelength range. 
The average seeing in the $J$ wavelength range was 1\farcs8, and the airmass was 1.3. 
In the $H$ wavelength range, the typical seeing was 1\farcs1 and the airmass was again about 1.3.
The data were spectroscopically calibrated with a standard star for flux calibration, which was observed during the same nights
as the science observations.
This was done for all observations setups, centering the telluric stars in the bluest and reddest slits of each mask. 
The uncertainty on the wavelength calibration, derived from the sky lines, was found to be $\sim$1~\AA, and it was consistent between the two 
gratings.

\section{Data reduction}
\label{sec:reduction}
Data reduction was made with the LUCI~1 spectroscopic reduction pipeline 
(pandora.lreducer\footnote{More details can be found at http://lbt-spectro.iasf-milano.inaf.it/lreducerInfo/.}). 
It consists of a set of tasks, based on VIPGI \citep[VIMOS Interactive Pipeline and Graphical Interface,][]{scodeggio05} recipes, which work in Linux systems
and are written in \textsc{c} and \textsc{python}.
Details on the reduction process are given in M12.
For the self-consistency of this paper, we here recall the main steps of the data reduction.
The preliminary steps of the pipeline work flow concern the creation of master calibration frames.
When all calibration frames were available, cosmic rays and bad pixels were removed, and dark and flat-field corrections 
were applied; then the spectra were extracted.
In this step, slit curvatures were removed and slits were extracted and wavelength calibrated.
Sky subtraction was made on 2D-extracted, wavelength-calibrated spectra. 
Further sky subtraction residuals were removed by subtracting the median computed along the spatial direction for each column of the spectrum. 
 
To detect the emission lines, we identified them in the 2D spectra. When we found an emission line in the 2D spectra, we extracted the single 1D spectra from the 2D stacked frame by coadding for each object all the rows containing the emission line features. This is 
shown for the confirmed protocluster members in the $J$ wavelength range in Fig.~\ref{fig:1DJ}.

Emission-line fluxes were measured with the task SPLOT of IRAF\footnote{IRAF is the Image Reduction and Analysis Facility. 
IRAF is written and supported by the National Optical Astronomy Observatories (NOAO) in Tucson, Arizona. 
NOAO is operated by the Association of Universities for Research in Astronomy (AURA), Inc. under cooperative agreement with the National Science Foundation.} 
in the 1D spectra.
We derived the redshifts with a Gaussian fit of the brightest emission line for each target, which allowed us to find the central wavelength
of the line. 
We derived uncertainties on the spectroscopic redshifts by adding in quadrature the fitting uncertainties
of all emission lines observed for a given source
and the uncertainty in the wavelength calibration. 

The signal-to-noise ratio (S/N) for the presumed [O~{\sc iii}]$\lambda$~5007~\AA\ emission lines of the newly identified galaxies (from CMC~2 to CMC~9) 
ranges from about 3 to approximately 8. 
To verify our identifications, it might be useful to detect the other line of the doublet of the [O~{\sc iii}], the [O~{\sc iii}]$\lambda$~4959~\AA, 
which has a flux of about 1/3 with respect to the [O~{\sc iii}]$\lambda$~5007~\AA. 
However, given the S/N of  the [O~{\sc iii}]$\lambda$~5007~\AA\ emission lines, the [O~{\sc iii}]$\lambda$~4959~\AA\ emission lines are expected 
to have an S/N from 1 to less than 3, which means that these lines will be very hard to detect.
In order to check the reliability of the redshift determination for these newly identified galaxies, we produced their stacked spectrum, 
as discussed in Sect.~\ref{sec:gal-z}.

Following G10 and M12, we assigned a 
quality  flag `A' or `B' to the measured redshifts: 
`A' indicates a secure spectroscopic redshift based on at least two spectral features, while `B' indicates high-level confidence in the
spectroscopic redshift based on a single spectral feature, almost always [O {\sc iii}]$\lambda$~5007~\AA. 
We added the flag `c' for the galaxies that are members and possible members of the protocluster (see Sect.~\ref{sec:results} for the definition of protocluster members). 
The redshift determinations with quality flag `A' correspond always to galaxies belonging to the protocluster, and they therefore also have a flag `c'.
The redshift determinations with quality flag `B' can either correspond to galaxies belonging to the protocluster (with flag `c') or to separate galaxies (without flag `c').   

Since the $H$ wavelength range part of the spectra was observed with a very short exposure time, we only detected the most intense emission lines of the four brightest objects: 
the radio galaxy, AGN1110, AGN1317, and AGN1354.
For AGN1110 we identified  only the H$\alpha$ emission line very close to a sky emission line in the $H$ wavelength range, while in the $J$ wavelength range 
we only detected the continuum spectrum.
The galaxy AGN1110 might have a spectrum with features similar to those of the central radio galaxy 7C~1756+6520, where only H$\alpha$ 
and H$\beta$ lines have been detected. 
In the case of AGN1110, the S/N is much lower, and thus H$\beta$ is below the detection limit. 
For this latter reason, Fig.~\ref{fig:1DJ} shows the 1D spectrum in the $J$ wavelength range but does not contain 
AGN1110.

\section{Results on the galaxy population of the protocluster}
\label{sec:results}  
In this section we present and discuss the results obtained from our LBT observations and their combination  
with other data that are available in the literature on the galaxy population that we identified in the protocluster around 
radio galaxy 7C~1756+6520.

\subsection{Emission-line identification and galaxy redshifts}
\label{sec:gal-z} 
The main results on emission-line identification, galaxy redshifts, and emission-line fluxes 
are collected in Tables~\ref{tab:z} and \ref{tab:fluxes}.
In Table~\ref{tab:z} we report the identification name for each galaxy (from G10 when available, otherwise they are identified 
with the name CMC [Casasola, Magrini, Combes, from the names of the first three authors of this work] followed by a number),
the J2000 RA and Dec. coordinates, the derived spectroscopic redshift with the flags as defined in Sect.~\ref{sec:reduction}, 
emission lines used to derive the redshift, the spectroscopic redshift from G10, 
and the corresponding emission lines we used to derive the redshift, when available, 
and the spectroscopic classification, AGN versus galaxy, based on spectra acquired so far for these galaxies.
The galaxies taken from G10 and M12 that we list in Table~\ref{tab:z} are all members of the protocluster, 
although their redshift determinations are quoted without the flag c. 
This flag was defined in this paper because our observations comprise both members and non-members of the protocluster. 
In Table~\ref{tab:z} we preferred to quote the original flags assigned by G10 and M12. 
Table~\ref{tab:fluxes} collects the observing wavelengths, the measured emission-line fluxes, and the FWHM for each detected source.    

The greatest part of the redshift identification is based on the detection of the H$\alpha$, H$\beta$, and [O~{\sc iii}]$\lambda$~5007~\AA\ emission lines.
For the four galaxies that have previously been confirmed as protocluster members (the radio galaxy, AGN1110, AGN1317, and AGN1354), we derived the redshift by detecting 
emission lines that have never before been detected for these sources. 
For the central radio galaxy 7C~1756+6520 we detected for the first time H$\alpha$ and H$\beta$ 
emission lines ($z = 1.4152 \pm 0.0009$, Ac).
This redshift determination is perfectly consistent within the errors with that of G10 for radio galaxy 7C~1756+6520 (see Table~\ref{tab:z}).  
Our observations confirm AGN1317 as a particularly interesting source: 
after the detection of [Ne~{\sc v}]$\lambda$~3426~\AA\ and [O~{\sc ii}]$\lambda$~3727~\AA\ by G10, 
we confirm the detection of M12  of very bright emission lines in 
H$\alpha$, [N~{\sc ii}]$\lambda\lambda$~6548, 6583~\AA, [S~{\sc ii}]$\lambda\lambda$~6716, 6731~\AA, 
H$\beta$, and [O~{\sc iii}]$\lambda\lambda$~4959, 5007~\AA. 
Based on four emission lines, we derive a redshift of $z = 1.4147 \pm 0.0010$ (Ac) for AGN1317.
This redshift is consistent within the errors with that found by G10 ($z = 1.4162 \pm 0.0005$, A),
based on two emission lines (see Table~\ref{tab:z}).  
For AGN1354, we detected the H$\alpha$, H$\beta$, and [N~{\sc ii}]$\lambda$~6583~\AA\ emission lines, 
which allowed us to unambiguously determine the redshift of this source ($z = 1.4177 \pm 0.0009$, Ac).
The redshift of AGN1354 has been derived by G10 by detecting only the Mg~{\sc ii}$\lambda$~2800~\AA\ emission line 
($z = 1.4153 \pm 0.0003$, B). 
We stress that our redshift determination for AGN1354 is based on more lines than were derived by G10.
For AGN1110, we detected for the first time the H$\alpha$ emission line, which it is quite remarkable because the exposure time in the $H$ wavelength range is short.
Based on this single emission line, we derived a redshift of $z = 1.4148 \pm 0.0001$ (Bc) for AGN1110, which is not consistent within the errors with (although similar to) the redshift determination of G10 ($z = 1.3935 \pm 0.0001$, B)
based on the single emission line of [O~{\sc ii}]$\lambda$~3727~\AA\ (see Table~\ref{tab:z}).

We also mention that in case of AGN, strong outflows can bias the redshift determination. 
However, only for AGN1317 has an outflow been detected so far (M12).  
The [O~{\sc iii}]$\lambda\lambda$~4959, 5007~\AA\ lines show a clear asymmetric profile with prominent blueshifted wings. This is 
a characteristic signature of outflows (see Fig.~6 in M12).

We identified a new protocluster member, CMC~1, by detecting H$\beta$ and [O~{\sc iii}]$\lambda$~5007~\AA\ emission lines.
We assigned a redshift of $1.4055 \pm 0.0047$ (Ac) to CMC 1, which is consistent with that of the central radio galaxy. 
We identified eight new possible protocluster members (CMC~2 -- 9) by assuming that the detected emission line is the [O~{\sc iii}]$\lambda$~5007~\AA\ 
for almost all cases and the H$\beta$ line for one galaxy (CMC~7).
These emission-line identifications correspond
to galaxies belonging to the redshift range of $z \sim 1.38 - 1.43$ (Bc).  
To maximize the S/N, we produced a rest-frame stacked spectrum by combining the spectra of the possible protocluster members 
in which we detected the [O~{\sc iii}]$\lambda$~5007~\AA\ line. 
This is shown in Fig.~\ref{fig:1DJ-stack}, where both the [O~{\sc iii}]$\lambda$~4959~\AA\ and H$\beta$ emission lines are detected. 
We also measured the ratio between the two lines of the [O~{\sc iii}] doublet and found that it is consistent with the
theoretical value of $\sim$3 \citep{osterbrock06}. 
This confirms the reliably of the line identification and that these galaxies belong to the protocluster.  

Finally, we detected one emission line for another two unknown sources (CMC~10 and CMC~11).
We identified these lines as the H$\beta$ line for CMC~10 and CMC~11.
If this emission-line assignation is correct, these sources would be at redshift $z \sim$1.37 (B).

\begin{sidewaystable*}
\caption[]{Our observations with detection (top part of the Table, tagged with ``Our observations'') and proto-cluster members information given from the literature and used in our analysis 
(bottom part of the Table, tagged with ``Proto-cluster members from the literature'').}
  \centering
  \begin{adjustbox}{max width=\textwidth}
\begin{tabular}{lllllllllll}
\hline
\hline
\textbf{Our observations}\\
\hline
ID      & Name  & RA (J2000)                                    & Dec (J2000)   & Redshift        $^{(a)}$ (this work) & Lines (this work) & Redshift$^{(a)}$ (G10) & Lines (G10) & Notes$^{(b)}$ \\
        &               & [$^{\rm h}$ $^{\rm m}$ $^{\rm s}$]    & [$^{\circ}$ $^{\prime}$ $^{\prime\prime}$]\\
\hline
\underline{Protocluster members} \\    
slit1                           & AGN1110              & 17:56:52.56   & 65:16:56.65     & 1.4148 $\pm$ 0.0001 (Bc)              & H$\alpha$                                             & 1.3935 $\pm$ 0.0012 (B) & [O~{\sc ii}]3727 & AGN \\
slit2                           & AGN1317              & 17:56:55.76   & 65:19:07.00     & 1.4147 $\pm$ 0.0010 (Ac)              & H$\alpha$, H$\beta$, [O~{\sc iii}]4959,5007, [N~{\sc ii}]6549,6583 & 1.4162 $\pm$ 0.0005 (A) & [Ne~{\sc v}]3426, [O~{\sc ii}]3727 & AGN  \\
slit3a                          & AGN1354              & 17:57:05.28   & 65:19:53.62     & 1.4177 $\pm$ 0.0009 (Ac)              & H$\alpha$, H$\beta$, [N~{\sc ii}]6583 & 1.4153 $\pm$ 0.0003 (B) & Mg~{\sc ii} 2800 & AGN\\
slit3b                          & 7C1756~+~6520 & 17:57:05.48   & 65:19:53.75   & 1.4152 $\pm$ 0.0009 (Ac)                & H$\alpha$, H$\beta$ & 1.4150 $\pm$ 0.0001 (A) & [Ne~{\sc v}]3426, [O~{\sc ii}]3727, [Ne~{\sc iii}]3869 & H$z$RG, AGN  \\
slit5                           & CMC~1                 & 17:56:58.12   & 65:18:18.09     & 1.4055 $\pm$ 0.0047 (Ac)              & H$\beta$, [O~{\sc iii}]5007  \\
\hline
\underline{Possible protocluster members} \\
slit4                           & CMC~2                 & 17:56:56.84   & 65:18:10.29     & 1.4079 $\pm$ 0.0004 (Bc)              & [O~{\sc iii}]5007 & & & Galaxy \\         
slit6                           & CMC~3                 & 17:56:53.19   & 65:17:47.93     & 1.4067 $\pm$ 0.0002 (Bc)              & [O~{\sc iii}]5007 & & & Galaxy \\
slit7                           & CMC~4                 & 17:56:51.86   & 65:17:38.91     & 1.3845 $\pm$ 0.0001 (Bc)              & [O~{\sc iii}]5007 & & & Galaxy \\
slit8                           & CMC~5                 & 17:56:46.60   & 65:17:49.40     & 1.4039 $\pm$ 0.0002 (Bc)              & [O~{\sc iii}]5007 & & & Galaxy \\
slit12                          & CMC~6                 & 17:57:08.25   & 65:18:49.88     & 1.3950 $\pm$ 0.0001 (Bc)              & [O~{\sc iii}]5007 & & & Galaxy \\
slit14                          & CMC~7                 & 17:57:00.34   & 65:19:09.47     &  1.4233 $\pm$ 0.0001 (Bc)             & H$\beta$           & & & Galaxy \\
slit16                          & CMC~8                 & 17:56:48.46   & 65:17:05.60     & 1.3999 $\pm$ 0.0001 (Bc)              & [O~{\sc iii}]5007 & & & Galaxy \\
slit17                          & CMC~9                 & 17:57:09.17   & 65:19:52.75     & 1.4336 $\pm$ 0.0001 (Bc)              & [O~{\sc iii}]5007 & & & Galaxy \\
\hline
\underline{Non protocluster members} \\
slit9                           & CMC~10                        & 17:57:08.56   & 65:19:00.29     & 1.3683 $\pm$ 0.0001 (B)               & H$\beta$      & & & Galaxy \\                                                                                                                           
slit18                          & CMC~11                        & 17:56:57.60   & 65:18:34.43     & 1.3673 $\pm$ 0.0001 (B)               & H$\beta$      & & & Galaxy \\
\hline 
\hline
\textbf{Protocluster members from the literature}\\
\hline 
ID      & Name  & RA (J2000)                                    & Dec (J2000)   & Redshift$^{(d)}$          & Lines $^{(c)}$  & Ref. \\
        &               & [$^{\rm h}$ $^{\rm m}$ $^{\rm s}$]    & [$^{\circ}$ $^{\prime}$ $^{\prime\prime}$]\\
\hline 
Cl 1756.2               & $sBzK$.6355   & 17:57:59.55 & 65:18:00.64 & $1.4020 \pm 0.0001$ (A) & [O~{\sc ii}] & G10 & & Galaxy   \\
Cl 1756.3               & $sBzK$.9622   & 17:57:21.75 & 65:23:05.04 & $1.4064 \pm 0.0005$ (A) & [O~{\sc ii}] & G10 & & Galaxy \\
Cl 1756.4               & $sBzK$.7556 $^{(c)}$  & 17:57:46.54 & 65:20:00.48 & $1.4081 \pm 0.0007$ (A) & Broad Mg~{\sc ii}, [O~{\sc ii}] & G10 & & AGN \\
Cl 1756.5               & $pBzK$.5858   & 17:57:44.40 & 65:17:14.30 & $1.4110 \pm 0.0010$ (B) & [O~{\sc ii}] & G10 & & Galaxy  \\
Cl 1756.6               & serendip.1            & 17:56:57.67 & 65:18:49.45 & $1.4150 \pm 0.0005$ (A) & [O~{\sc ii}] & G10 & & Galaxy  \\
Cl 1756.8               & serendip.2            & 17:57:25.00 & 65:19:04.83 & $1.4157 \pm 0.0010$ (A) & [O~{\sc ii}] & G10 & & Galaxy  \\
Cl 1756.9               & $sBzK$.6997   & 17:57:24.43 & 65:19:03.87 & $1.4157 \pm 0.0006$ (A) & [O~{\sc ii}] & G10 & & Galaxy  \\
Cl 1756.11      & $sBzK$.7624   & 17:56:33.16 & 65:20:04.46 & $1.4236 \pm 0.0001$ (A) & Fe~{\sc ii}+Mg~{\sc ii} absn, [O~{\sc ii}] & G10 & & Galaxy   \\
Cl 1756.12      & $pBzK$.7523   & 17:57:05.04 & 65:19:54.50 & $1.4244 \pm 0.0004$ (A) & [O~{\sc ii}] & G10 & & Galaxy  \\
Cl 1756.13      & $sBzK$.5860 $^{(c)}$  & 17:57:35.34 & 65:17:14.39 & $1.4268 \pm 0.0005$ (A) & Broad Mg~{\sc ii}, [O~{\sc ii}] & G10 & & AGN   \\
Cl 1756.14      & $sBzK$.5699   & 17:57:44.06 & 65:16:57.11 & $1.4274 \pm 0.0004$ (A) & [O~{\sc ii}] & G10 & & Galaxy   \\
Cl 1756.15      & $pBzK$.10235  & 17:57:25.20 & 65:23:58.19 & $1.4277 \pm 0.0010$ (B) & [O~{\sc ii}] & G10 & & Galaxy  \\
Cl 1756.16      & $sBzK$.4449   & 17:57:47.40 & 65:14:52.17 & $1.4326 \pm 0.0002$ (A) & [O~{\sc ii}] & G10 & & Galaxy  \\
Cl 1756.17      & $sBzK$.7625   & 17:57:14.41 & 65:20:02.40 & $1.4366 \pm 0.0001$ (A) & [O~{\sc ii}] & G10 & & Galaxy  \\
Cl 1756.18      & AGN.1206              & 17:57:13.08 & 65:19:08.37 & $1.4371 \pm 0.0002$ (A) & [O~{\sc ii}] & G10 & & AGN  \\
Cl 1756.19      & $sBzK$.7208   & 17:57:18.31 & 65:19:24.94 & $1.4374 \pm 0.0002$ (A) & [O~{\sc ii}] & G10 & & Galaxy  \\
Cl 1756.20      & serendip.3            & 17:57:20.76 & 65:19:39.14 & $1.4379 \pm 0.0007$ (A) & [O~{\sc ii}] & G10 & & Galaxy  \\
MSC2            &                               & 17:57:10.16 & 65:19:28.09 & $1.4556 \pm 0.0005$ (A) & H$\alpha$, [N~{\sc ii}] & M12 & & Galaxy \\
\hline 
\hline
\end{tabular}
\end{adjustbox}
\label{tab:z}
\tablefoot{
\tablefoottext{a}{
The assignment of the quality flags to the redshifts is the same as was adopted by G10 and M12.
We also adopted the flag `c' for the redshifts corresponding to galaxies belonging to the protocluster.
The flag `c' is not present for the redshifts from G10 and M12, although 
the galaxies from G10 and M12 are all defined as protocluster members.     
}
\tablefoottext{b}{
Spectroscopic classification, AGN vs. galaxy, based on spectra acquired so far.
}
\tablefoottext{c}{
$sBzK$.7556 and $sBzK$.5860 from G10, both targeted as $sBzK$ galaxies, are classified as AGN 
because their spectra show AGN signatures such as strong, broad Mg~{\sc ii} emission lines.  
}
\tablefoottext{d}{
Redshifts derived from G10 based on [O~{\sc ii}] and with quality flag `A' mean that  
the [O~{\sc ii}]$\lambda$~3727~\AA\ emission has been clearly detected as a doublet.
}
}
\end{sidewaystable*}

\begin{table*}
\centering
\caption[]{Measured emission-line fluxes.}
\begin{tabular}{lllll}
\hline
\hline
Name            &Lines                  &  Wavelength   & Flux                                  & FWHM \\
                        &                               &[\AA]          & [erg~cm$^{-2}$~s$^{-1}$]        & [\AA] \\
\hline
AGN1110                                & H$\alpha$                             & 15848.6$\pm$0.6                 & 1.4$\pm$0.2E--16              & 12$\pm$1\\
AGN1317                                & H$\beta$ n+b $^{2}$           & 11740.2$\pm$0.6         & 4.2$\pm$0.5E--16              & 93$\pm$1    \\
                                        & [O~{\sc iii}]4959 n+b $^{2}$  & 11971$\pm$1     .0              & 1.15$\pm$0.5E--16     & 28$\pm$3 \\
                                        & [O~{\sc iii}]5007 b $^{2}$    & 12084.8$\pm$0.6        & 1.7$\pm$0.3E--16          &39$\pm$0.6\\
                                        & [O~{\sc iii}]5007 n $^{2}$    &12094.6$\pm$0.6           & 5.8$\pm$0.3E--16          & 23$\pm$1 \\
AGN1317         (M12) $^{1}$   & H$\alpha$ n $^{2}$            &15861.0$\pm$0.6                 &2.8$\pm$0.1E--16       &9$\pm$2  \\
                                        &[N~{\sc ii}]6548                       &15812.7$\pm$0.6                 &5.2$\pm$1.0E--17       &9$\pm$2 \\
                                        &[N~{\sc ii}]6583                       &15906.9$\pm$0.6                &9.3$\pm$1.0E--17         &9$\pm$2 \\
                                        &H$\alpha$ b $^{2}$             &15851.4$\pm$0.6                 &9.7$\pm$0.3E--16       &121$\pm$1\\ 
                                        &[S~{\sc ii}]6717                       &16228.3$\pm$0.6                 &4.4$\pm$1.0E--17       &9$\pm$2  \\
                                        &[S~{\sc ii}]6731                       &16264.5$\pm$0.6                 &2.0$\pm$1.0E--17       &9$\pm$2 \\                     
AGN1354                                & H$\beta$              & 11753.1$\pm$0.9               &1.1$\pm$0.1E--16        & 34$\pm$2\\
                                        & H$\alpha$             & 15859.0$\pm$0.7         & ---   $^3$                    & --- \\
                                        & [N~{\sc ii}]6583      & 15919.1$\pm$1.4               & ---    $^3$                    & --- \\
7C1756+6520                     & H$\beta$              & 11737.3$\pm$0.7         & 2.6$\pm$0.3E--16      & 27$\pm$1      \\
                                        & H$\alpha$             & 15850.06                      &  ---    $^3$    &  ---  $^3$ \\
CMC~1                           & H$\beta$              & 11710.0$\pm$0.7               & 2.6$\pm$1E--18 & 12$\pm$2 \\
                                        & [O~{\sc iii}]5007     & 12027.4$\pm$0.2               & 2.1$\pm$0.1E--16        & 11$\pm$1 \\
CMC~2                           & [O~{\sc iii}]5007     & 12056.0$\pm$2.0                 & 8.0$\pm$1E--17        & 40$\pm$5\\            
CMC~3                           & [O~{\sc iii}]5007     & 12050.0$\pm$1.0               & 5.0$\pm$1E--17 & 18$\pm$2      \\
CMC~4                           & [O~{\sc iii}]5007     & 11939.0$\pm$0.5               & 4.5$\pm$1E--17 & 10$\pm$1    \\
CMC~5                           & [O~{\sc iii}]5007     & 12037.0$\pm$1.0               & 4.5$\pm$1E--17 & 10$\pm$1   \\
CMC~6                           & [O~{\sc iii}]5007     & 11991.3$\pm$0.5               & 3.0$\pm$1E--17 & 13$\pm$2    \\
CMC~7                           & H$\beta$              & 11780.5$\pm$0.5               & 2.6$\pm$1E--17  & 10$\pm$1   \\
CMC~8                           & [O~{\sc iii}]5007     & 12016.0$\pm$0.4               & 6.0$\pm$1E--17  & 13$\pm$2  \\
CMC~9                           & [O~{\sc iii}]5007     & 12184.6$\pm$0.2               & 7.2$\pm$1E--17 & 15$\pm$2     \\
CMC~10                          & H$\beta$              & 11513.0$\pm$1.0               & 8.0$\pm$1E--17  & 15$\pm$2     \\                                                                                                                               
CMC~11                          & H$\beta$              & 11508.4$\pm$0.1              & 1.15$\pm$0.1E--16 & 10$\pm$1   \\
\hline 
\hline
\end{tabular}
\label{tab:fluxes}
\tablefoot{ 
$^{1}$ Emission-lines fluxes taken by M12 (only for AGN1317) because of the short observing time in the $H$ wavelength range of our observations; 
$^{2}$ n = narrow component, b = broad component, n+b = narrow+broad components (when it has not been possible to separate the two components); 
$^{3}$ H$\alpha$ and [N~{\sc ii}] lines superposed to sky-emission.
}
\end{table*}

\begin{figure*}
\centering
\includegraphics[width=0.95\textwidth]{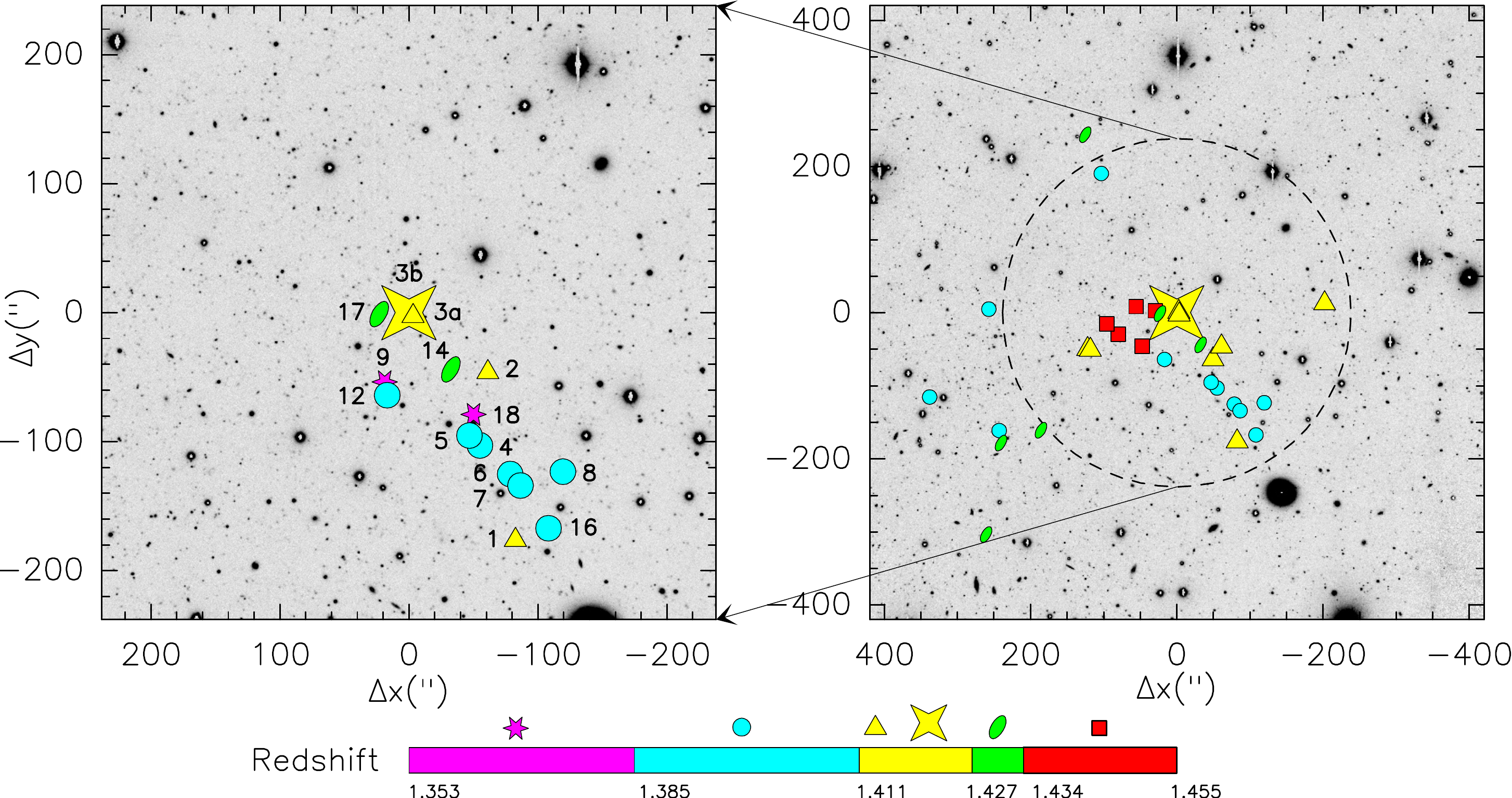}
\caption{\textit{Left panel:}
Galaxies we spectroscopically detected in the field of the overdensity around radio galaxy 7C~1756+6520 plotted in the $B$-band image (NOAO).
They comprise protocluster members, possible protocluster members, and non protocluster members (see Table~\ref{tab:z}).    
ID numbers, corresponding to the slit,  are provided for each source (see Table~\ref{tab:z}).
The five colors and symbols indicate different redshift ranges: magenta stars $z < 1.385$, light blue circles $1.385 \leq z \leq 1.411$, 
yellow triangles $1.411 < z < 1.427$, green ellipses $1.427 \leq z \leq 1.434$, and red squares (in the right panel) $z > 1.434$.   
The central radio galaxy 7C~1756+6520 is the large yellow star in the center.
North is up and east to the left. 
The FoV is about 480\arcsec~$\times$~480\arcsec, centered on the central radio galaxy. 
1\arcsec\ corresponds to $\sim$8.4~kpc.
\textit{Right panel:} 
All protocluster and possible protocluster members with spectroscopically confirmed redshift from this work, G10, and M12. 
Symbols and colors are the same as in the left panel.
The FoV is 840\arcsec~$\times$~840\arcsec.  
The dashed circles indicate a distance of 2~Mpc from the central radio galaxy.
}
\label{fig:cluster}
\end{figure*}

\subsection{Protocluster members and their spatial and redshift distribution within the structure}
\label{sec:gal-distr} 
The left panel of Fig.~\ref{fig:cluster} shows the spatial distribution of the 15 sources we detected in the field of the protocluster (see Table~\ref{tab:z}).
There is no clear and consensual definition of a galaxy cluster.
\citet{eisenhardt08} defined a $z > 1$ cluster as spectroscopically confirmed if it
contains at least five galaxies within a radius of 2~Mpc  whose spectroscopic redshifts
match to within $\pm2000\,(1 + <z>)$~km~s$^{-1}$.
According to this definition, 13 of the 15 objects we observed would be part of the protocluster.
Since 4 of these 13 galaxies (the radio galaxy and the 3 AGN) have previously been classified as protocluster members, we would add 9 new protocluster 
members.
For these 9 proto-cluster members, we give a secure spectroscopic redshift for CMC~1 (classified as `Protocluster members' in Table~\ref{tab:z}),
while we attribute a redshift based on only one emission line to the other 8 sources
(CMC~2 -- 9, classified as `Possible protocluster members' in Table~\ref{tab:z}).
However, the spectroscopically confirmed high-redshift galaxy clusters in \citet{eisenhardt08} and in other works 
\citep[e.g.,][]{hilton07,hayashi10,hayashi14} usually show much narrower velocity distributions ($\sim$1000 -- 2000~km~s$^{-1}$).
The most part of the new detected galaxies also satisfies these more stringent conditions.
To verify this last definition of cluster members, 
the peculiar velocity of a galaxy with redshift $z$ in the rest-frame of a cluster with redshift $z_{\rm {cl}}$ is defined as
\begin{eqnarray}
v_{\rm {pec}}^{\rm {rest}} = c(z - z_{\rm {cl}})/(1 + z_{\rm {cl}}) & \mbox{(for }v_{\rm {pec}}^{\rm {rest}}\ll c)
\label{eq:vpec}
\end{eqnarray}
\noindent
\citep[e.g.,][]{milvang-jensen08}.
Based on Eq.~(\ref{eq:vpec}) and assuming that the protocluster redshift is that of the radio galaxy, 
$z_{\rm {cl}} = 1.4152$, galaxies CMC~1, CMC~2, CMC~3, and CMC~7  have peculiar velocity offset of 
$\sim$900--1200~km~s$^{-1}$ with respect to the radio galaxy, and that of galaxies CMC~5, CMC~6, CMC~8 and CMC~9 is  
$\sim$1400--2500~km~s$^{-1}$ with respect to the radio galaxy. 
Galaxy CMC~4 instead has a peculiar velocity offset of $\sim$3800~km~s$^{-1}$ with respect to the radio galaxy.
Under the hypothesis that the emission line detected in galaxies CMC~10 and CMC~11 is the H$\beta$ line, 
these sources would have peculiar velocities offset by $\sim$$5800 - 5950$~km~s$^{-1}$ with respect to the central radio galaxy. 
Although these two galaxies fall in the field of the galaxy overdensity, they are not expected to be part of it 
(they are classified as `Non-protocluster members' in Table~\ref{tab:z}).
Additional spectroscopic observations are required to define the redshift of these sources, in particular, deeper $H$ wavelength range 
observations.

By merging the spectroscopically confirmed protocluster members (taking into account the `Possible protocluster members'
of Table~\ref{tab:z} as well) derived from this work, G10, and M12, 31 galaxies, including the central radio galaxy, are found
around redshift $1.4152 \pm 0.056$. This corresponds to peculiar velocities $\lesssim$5000~km~s$^{-1}$ 
with respect to the radio galaxy (or $\lesssim$3800~km~s$^{-1}$ excluding the galaxy MSC2).
For the protocluster members of G10 and M12 that we used in our analysis, we refer to the bottom part of Table~\ref{tab:z}.
For the protocluster members in common with G10 and M12, we adopted our redshift determinations.
This result is consistent with that of G10, who found 21 galaxies, including the central radio galaxy, at redshift 
$1.4156 \pm 0.025$. This corresponds to peculiar velocities $\lesssim$3000~km~s$^{-1}$ with respect to the radio galaxy.   

The right panel of Fig.~\ref{fig:cluster} shows all 31 galaxies that are spectroscopically confirmed members of the overdensity around radio galaxy 7C~1756+6520.   
To better compare this figure with Fig.~6 of G10, we adopted similar color and redshift ranges
(although we consider a larger redshift window and the redshift ranges are therefore slightly different).  
As described in Sect.~\ref{sec:cluster}, G10 have identified two subgroups in the protocluster: 
one that forms a large structure at the redshift of the radio galaxy, $z \sim 1.42$ (yellow triangles and large yellow stars), 
and another compact structure at $z \sim 1.44$ that  forms a subgroup offset by $\Delta v \sim 3000$~km~s$^{-1}$ 
and approximately 1\farcs5 east of the radio galaxy (red squares).
The right panel of Fig.~\ref{fig:cluster} seems to suggest that another large subgroup lies at $z \sim 1.40$ (light blue circles), 
which is similar to the subgroup at $z = 1.42$ in terms of spatial distribution.

\begin{figure}
\centering
\includegraphics[width=0.5\textwidth]{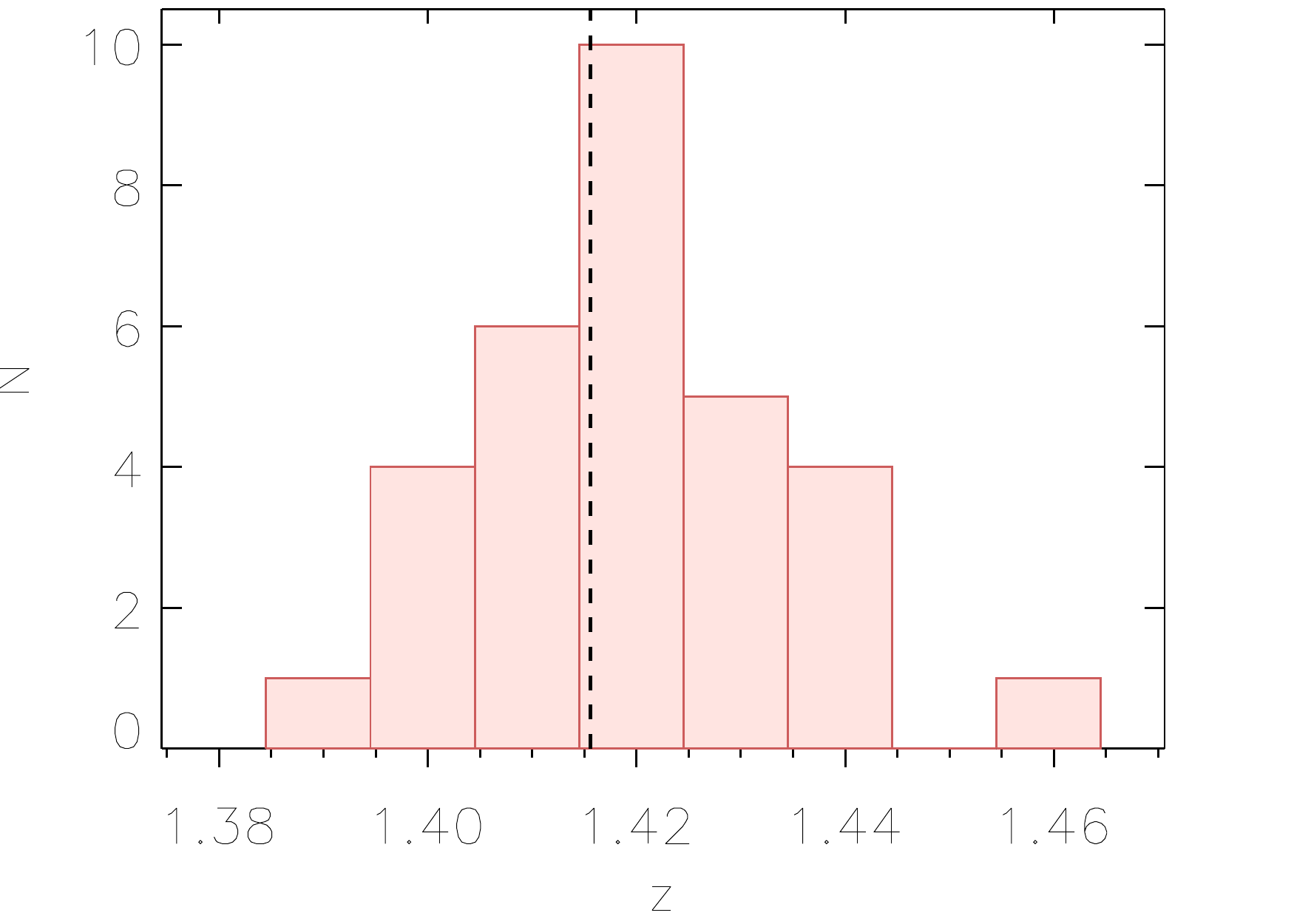}
\caption{Redshift histogram for the spectroscopically confirmed members and possible members  ($N = 31$) of the protocluster
around H$z$RG 7C~1756+6520 at $z = 1.4156$. 
The plotted redshifts come from this work, G10, and M12.
For the protocluster galaxies in common with G10 and M12, we adopted our redshift determinations.
Redshift bins are of 0.01 ($\sim$3000~km~s$^{-1}$ observed frame), 
corresponding to the standard deviation of the Gaussian function distribution.
The vertical dashed line marks the redshift of the central H$z$RG 7C~1756+6520.  
}
\label{fig:histo}
\end{figure}

A way to better characterize the redshift distribution of protocluster members within the overdensity is shown in Fig.~\ref{fig:histo}: 
it displays the redshift histogram obtained by plotting all 31 spectroscopically confirmed protocluster and possible protocluster
members.
Based on the available statistics, the redshift distribution is dominated by one main peak in redshift space at $ z = 1.42$ 
that can be described by a Gaussian function with standard deviation of $\sigma = 0.01$. 
The Gaussian distribution probability in the range of redshift values [$-\sigma$, $+\sigma$] is $52\%$, in the range
[$-2\sigma$, $+2\sigma$] it is $84\%,$  and in the range [$-3\sigma$, $+3\sigma$] it is $97\%$. 
This result agrees with the typical redshift distribution within high-redshift galaxy clusters
\citep[we refer, e.g., to][and following publications in the context of the ESO Distant Cluster Survey, EDisCS]{halliday04}.    
Substructures in the redshift histogram are sometimes found in high-redshift galaxy clusters
\citep[e.g.,  at $z = 0.75,$ Cl1054--12 shows an evident secondary peak, and at $z = 0.54,$ Cl1232 shows a substructure in their redshift histograms,][]{halliday04}.
However,  it is not immediately obvious that these substructures  are spatially offset from each other on the sky.

That protocluster members lie even at large scale with respect to 7C~1756+6520 is consistent with 
 a filamentary nature of overdensities and that they extend beyond the projected dimension of 
$\sim$2 Mpc at intermediate redshift (see, e.g., the galaxy overdensity at $z \sim 0.4$ studied by Zappacosta et~al.~2002 and 
Mannucci~et~al.~2009) and at high redshift \citep[see, e.g., the protocluster around PKS~$1138-262$ at $z \sim 2.16$:][]{pentericci98,carilli02,croft05}. 
This might suggest that we are seeing galaxies located in a filament 
of the large-scale structure associated with the protocluster around 7C~1756+6520.

\subsection{Cluster velocity dispersion}
\label{sec:vel-disp} 
Following the method of \citet{beers90} for estimating the kinematical properties of galaxy clusters
for a few members,
we found that the velocity dispersion of the sources within 2~Mpc from the radio galaxy 
(i.e., those classified as `Protocluster members' and `Possible protocluster members' in Table~\ref{tab:z})
is $\sigma_{\rm {cl}} \sim 1700 \pm 140$~km~s$^{-1}$.
This value is consistent within 3$\sigma$ with the velocity dispersion found by G10 ($\sigma_{\rm {cl}} \sim 1270 \pm 180$~km~s$^{-1}$) when we
adopt the same method for the cluster members that they identified in almost the same region as we considered.    
Despite this agreement, these values of the velocity dispersion are only indicative because of 
the small numbers of galaxies used by us and G10 in deriving the velocity dispersion. 

Measurements of cluster velocity dispersions are often used to estimate cluster masses \citep[e.g.,][]{fisher98,tran99,borgani99,lubin02}. 
Early measurements of velocity dispersions \citep[e.g.,][]{danese80} have typically assumed a Gaussian velocity distribution. 
As datasets have increased in quality and size, significant deviations from Gaussian behavior have often been found 
\citep[e.g.,][]{zabludoff93,fisher98}.
These are likely associated with the substantial structural irregularities (usually termed substructures)
that are observed in many clusters \citep[][]{geller82,dressler88}. 
Any statistics used to estimate the cluster velocity dispersion should therefore be robust against outliers and against variations in the shape of the underlying
velocity distribution \citep[][]{beers90}. 
However, since the overdensity around radio galaxy 7C~1756+6520 has a velocity dispersion exceeding 1000~km~s$^{-1}$
and it is unlikely to be fully virialized, its velocity dispersion may therefore not provide a robust estimate of its
mass.
For this reason, we do not provide an estimate of the protocluster mass from the velocity dispersion.

\subsection{Dressler-Shectman test}
\label{sec:dressler}
In order to search for substructures in three-dimensional space (which were only qualitatively discussed in Sect.~\ref{sec:gal-distr} 
and in the previous works on this protocluster), we combined velocity and positional information by computing the robust statistics 
devised by \citet{dressler88}.
Starting from a list of cluster members with measured positions and velocities, the listed 10 nearest neighbors on the sky are determined.
The local mean velocity and velocity dispersion are computed from this sample of 11 galaxies.
These quantities, defined for each galaxy in the list, are then compared to the global cluster mean velocity and velocity dispersion using the parameter $\delta,  $ which is defined
as\begin{eqnarray}
\rm {\delta}^2 &=& {\frac{11}{\sigma_{\rm {cl}}^2}} \times \left[ (\overline{v}_{\rm {local}}- \overline{v}_{\rm {cl}})^2 + (\sigma_{\rm {local}}- \sigma_{\rm {cl}})^2 \right]
\label{eq:dressler}
,\end{eqnarray}
\noindent
where $\overline{v}_{\rm {cl}}$ and and $\sigma_{\rm {cl}}$ are the cluster velocity 
and the cluster velocity dispersion, respectively,  while $\overline{v}_{\rm {local}}$ and $\sigma_{\rm {local}}$ are the locally defined quantities for each galaxy.

\citet{dressler88} also defined a cumulative deviation $\Delta$, which measures the amount of clumpiness in the cluster,
that is, the sum of  $\delta$ for $N_g$ cluster members. This is expressed as
\begin{eqnarray}
\rm {\Delta} &=& \sum_{i=1}^N \delta_{i}
\label{eq:delta}
.\end{eqnarray}
\noindent
The $\Delta$ statistics is similar to $\chi^2$, although \citet{dressler88} chose not to square the derivations 
before summation in order to de-emphasize the largest, rarest derivations. 
If the cluster velocity distribution is close to Gaussian 
and local variations are only random fluctuations, then $\Delta$ will be on the order of $N_g$.

We applied the Dressler-Shectman (DS) test starting from the 31 protocluster and possible protocluster members 
identified so far. 
This number is higher than the formal minimum number required to perform the test (11), 
and is slightly higher than the lower limit (26) considered by \citet{dressler88} to apply their method.  

\begin{figure}
\centering
\includegraphics[width=0.45\textwidth]{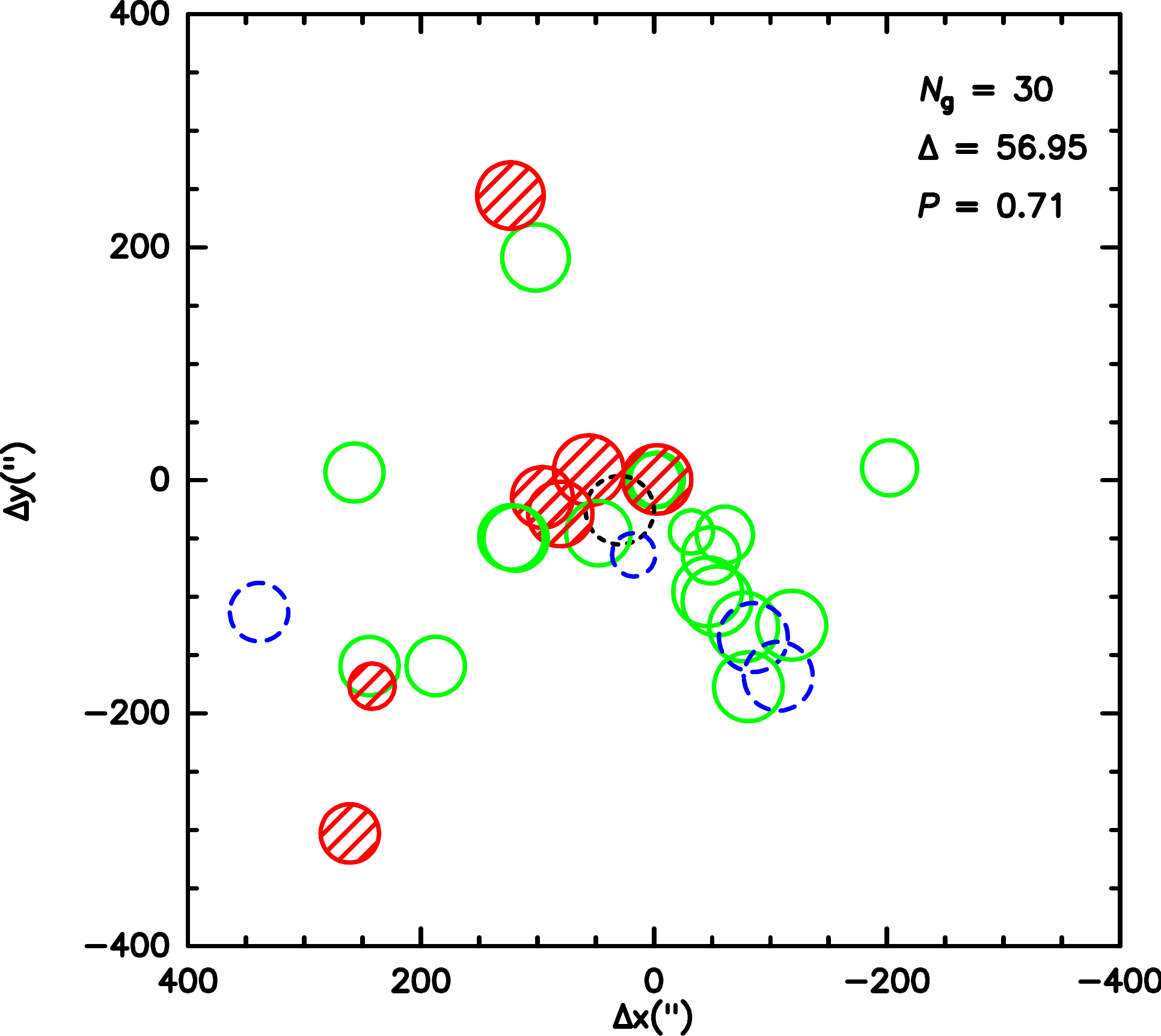}
\caption{Dressler-Shectman (DS) plot. 
It shows the $x$, $y$ location of the spectroscopically-confirmed proto-cluster members, centered on 7C~1756+6520. 
The DS test is applied to proto-cluster members with $z$ within the range 
[$-3\sigma_{\rm cl}$, $+3\sigma_{\rm cl}$] from $z_{\rm {cl}}$.
The radii of the plotted circles are proportional to $e^{\delta}/2$, where $\delta$ is 
the DS measurement of local deviation from the global velocity dispersion 
and mean recession velocity (cf. Eq.~\ref{eq:dressler}). 
Depending on in which bin $v_{\rm {pec}}^{\rm {rest}}$ falls, the circles are:
blue and dashed [$-3\sigma_{\rm cl}$, $-1\sigma_{\rm cl}$[, green [$-1\sigma_{\rm cl}$, $+1\sigma_{\rm cl}$], 
red and hashed ]$+1\sigma_{\rm cl}$, $+3\sigma_{\rm cl}$].   
The black and dotted circle is the galaxy MSC2, excluded from the DS test because 
it has a $v_{\rm {pec}}^{\rm {rest}}$ that does not fall in the range [$-3\sigma_{\rm cl}$, $+3\sigma_{\rm cl}$].
The number of proto-cluster members $N_{\rm g}$, 
the statistics $\Delta$, and the statistical significance $P$ are also given on the figure.
}
\label{fig:dressler}
\end{figure}

This technique does not allow a direct identification of galaxies belonging to a detected substructure.   
The position of substructures can be identified by plotting the distribution of the galaxies 
on the sky, however, using symbols whose size is proportional to the parameter $\delta$, thus quantifying the local deviation
from the global kinematics of the cluster.
This plot is shown in Fig.~\ref{fig:dressler}, where the size of the symbols is proportional to $e^{\delta/2}$.
The circles are colored according to the criteria adopted by \citet{milvang-jensen08},
who applied the DS test to 15 clusters in the context of the EDisCS survey. 
By considering protocluster members with $z$ within the range [$-3\sigma_{\rm cl}$, $+3\sigma_{\rm cl}$] 
from $z_{\rm {cl}}$ and depending on in which bin $v_{\rm {pec}}^{\rm {rest}}$ falls, 
the circles in Fig.~\ref{fig:dressler} are as follows: 
blue and dashed [$-3\sigma_{\rm cl}$, $-1\sigma_{\rm cl}$[, green and solid [$-1\sigma_{\rm cl}$, $+1\sigma_{\rm cl}$], 
and red and hashed ]$+1\sigma_{\rm cl}$, $+3\sigma_{\rm cl}$].   
The protocluster member MSC2 given from M12 does not fall in the range [$-3\sigma_{\rm cl}$, $+3\sigma_{\rm cl}$],
and for this reason, MSC2 is excluded from the DS test.
The applied DS test is based on $N_{\rm g} = 30$ (31 protocluster members minus MSC2), giving $\Delta = 56.95$.  
To first order, the many large circles in a given area should correspond
to a substructure, that is, a spatial and kinematical correlation
of separate structures. 
The visual inspection of Fig.~\ref{fig:dressler} does not seem to suggest the presence of substructures. 

The $\Delta$ statistic can be used to give a quantitative estimate of the significance of putative substructures. 
We computed a set of 1000 Monte Carlo simulations by randomly reshuffling the velocities of the protocluster members. 
This removes any significant 3D substructure.
The statistical significance of an observed substructure can then be quantified by
the fraction $P$ \citep[][]{dressler88} of the simulations that yield $\Delta$ values higher than the observed value, where a low
value of $P$ (i.e., very close to 0 in a range between 0 and 1) corresponds to a high significance.
For $N_{\rm g} = 30$ and $\Delta = 56.95$, we obtained a substructuring significance $P = 0.71$.
This high value of $P$ confirms that we are not able to define the presence of substructures within protocluster 
7C~1756+6520 on a statistical basis.
 
\citet{halliday04} tested five galaxy clusters and detected significant substructure in two clusters:
Cl1232.5--1250 at $z = 0.54$ and Cl1216.8-1201 at $z = 0.79$.
\citet{halliday04} identified substructures in galaxy clusters with $\sim$55--70 available spectroscopic members, 
while no substructures were found in clusters with $\sim$30--50 available spectroscopic members.
Out of the nine galaxy clusters, \citet{milvang-jensen08} detected significant substructures in two clusters: Cl1037.9--1243a at
$z = 0.43$ and Cl1354.2--1230 at $z = 0.76$.
\citet{milvang-jensen08} instead identified substructures with $\sim$20--45 available spectroscopic members.  
The fraction of EDisCS clusters with detected substructures is 29$\%$.
A similar percentage, 31$\%$, was found by \citet{solanes99} for a local ($z \lesssim 0.1$) sample of clusters from the ESO Nearby Abell Cluster Survey (ENACS).

\subsection{Phase-space diagram}
\label{sec:phase-space}
In order to derive information on the orbital histories of galaxies belonging to the overdensity 7C~1756+6520,
we produced a position versus velocity phase-space diagram 
\citep[e.g.,][]{Hernandez-Fernandez14,haines15,jaffe15,noble16,Yoon17,jaffe18}.
The path a galaxy takes through a cluster, and thus the exposure of the galaxy to different density environments, is indeed encoded in its orbital history.
Unfortunately, the orbital history is not directly observable as we are limited to a single projected snapshot in time. 
However, simulations have shown that 
phase-space diagrams (the member galaxy line-of-sight velocity relative to the cluster versus clustercentric radius)
can help to circumvent this problem, as it is sensitive to the time since galaxy infall
\citep[e.g.,][]{gill05,haines12,taranu14}.
Moreover, phase-space diagrams can be used to study cluster processes such as ram-pressure stripping 
\citep[e.g.,][]{Hernandez-Fernandez14,jaffe15,Yoon17,jaffe18} or tidal mass loss \citep[e.g.,][]{Rhee17}.

Using the Millennium Simulation from \citet{Springel05}, \citet{haines12} traced out galaxy accretion histories
for orbiting galaxies of 30 massive clusters as a function of phase space 
\citep[see Figure 3 in][]{haines12}.
These diagrams reveal the distinct trumpet-shaped (or chevronshaped) loci occupied by the older, virialized population with respect to the
more recently accreted infall population \citep[see also][]{haines15}.
Thus, a phase-space analysis for the environment can effectively account for distinctive
cluster populations and alleviate some of the projection effects
that bias the traditional probes for environment: clustercentric
radius and local density.

\begin{figure}
\centering
\includegraphics[width=0.5\textwidth]{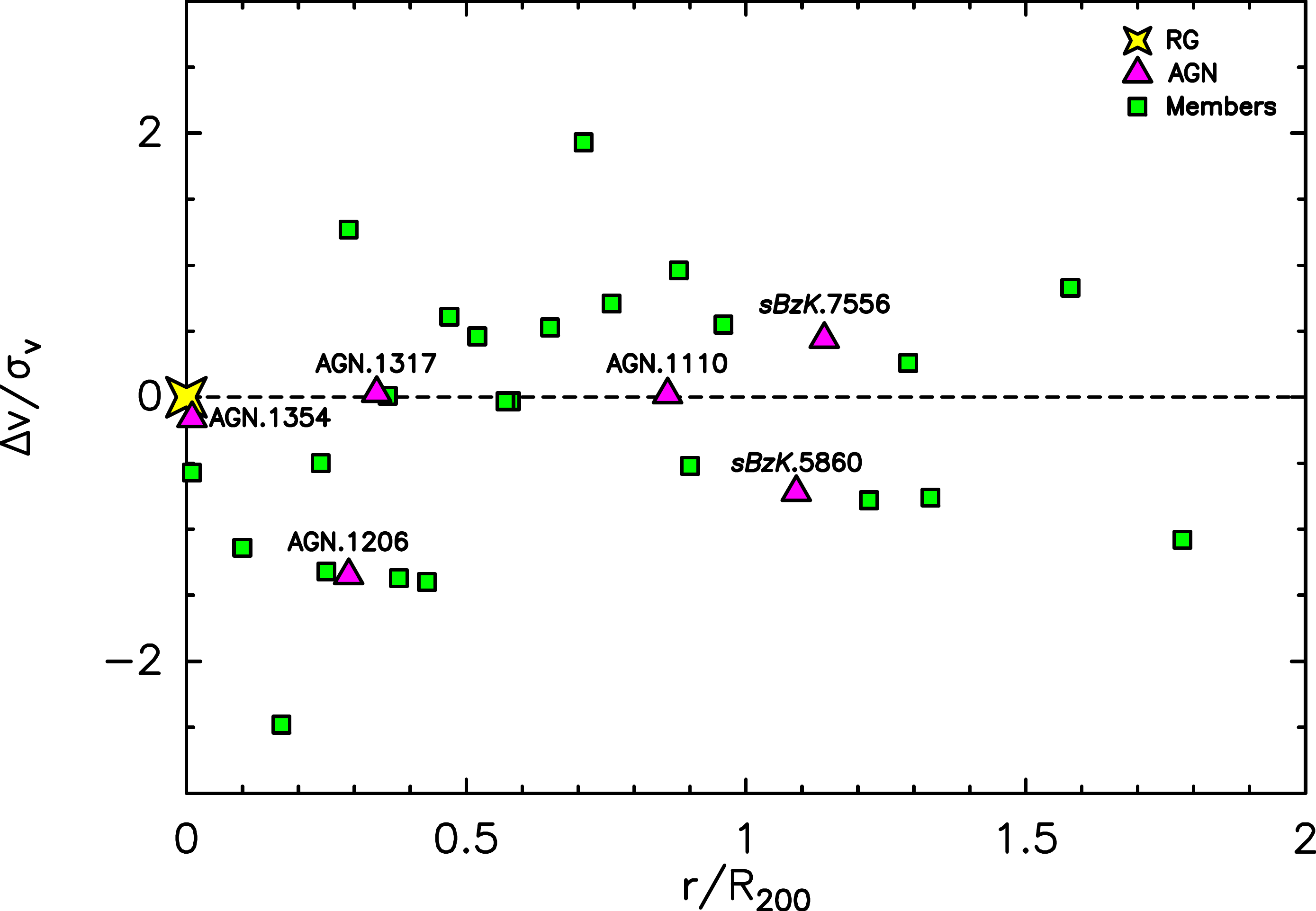}
\caption{
Phase-space diagram (line-of-sight velocity vs. clustercentric radius, normalized by the protocluster velocity dispersion and the radius $R_{200}$, respectively) for 
the spectroscopically identified 31 members of the overdensity 7C~1756+6520.
Different symbols refer to different types of protocluster members, as specified in the legend. 
RG refers to the central radio galaxy of the protocluster.
The names of the AGN are also reported. 
The dashed horizontal line is drawn at  $\Delta$v/$\sigma_{\rm v} = 0$,  corresponding to the systemic velocity of the protocluster.
}
\label{fig:space-phase}
\end{figure}

Figure~\ref{fig:space-phase} shows the projected phase-space diagram for the identified 31 members of protocluster 7C~1756+6520.
It displays line-of-sight velocity versus clustercentric radius, normalized by the cluster velocity dispersion, and the radius $R_{200}$, respectively.
By definition, $R_{200}$, which approximates the virial radius, is the radius inside which the density is 200 times the critical density.
The radius $R_{200}$ was computed from the cluster velocity dispersion as in \citet{poggianti06},
\begin{eqnarray}
R_{200}\, \rm{[Mpc]} &= & 1.73\,{\frac{\sigma_{\rm cl}}{1000\,\rm{km\,s^{-1}}}} {\frac{1}{\sqrt{\Omega_{\Lambda} + \Omega_0 (1 + z)^3}}}\,h^{-1}
\label{eq:r200}
,\end{eqnarray}
\noindent
where $\sigma_{\rm cl}$ has been computed in Sect.~\ref{sec:vel-disp} and the cosmological parameters have been defined in Sect.~\ref{sec:intro} 
\citep[see also][]{finn05}. 
Based on Eq.~(\ref{eq:r200}), we derived a value of $R_{200} = 1.89$~Mpc.

Figure~\ref{fig:space-phase} shows that three AGN (AGN1110, AGN1317, and AGN1354) of our sample 
lie at small projected protocluster centric radii ($ r/R_{200} < 1$) and have velocities that are remarkably similar to the systemic velocity of the protocluster
($\Delta$v/$\sigma_{\rm v} \simeq 0$).
Galaxies in the region in the phase-space diagram at low velocity and small radii have a high probability to be a ``virialized'' population that has been part 
of the protocluster for a long time \citep[e.g.,][]{Hernandez-Fernandez14,haines15,jaffe15,noble16,Rhee17,Yoon17,jaffe18}.
Although we recall that the projected velocities and radii are both lower limits to the 3D values of these quantities \citep[see Figure~6 in][]{jaffe18}, 
the position of AGN1110, AGN1317, and  AGN1354 in our diagram might suggest that these AGN and the central radio galaxy 
form a population that has been coexisting in the densest core region of this forming structure.
We performed the 2D Kolmogorov-Smirnov (KS) test \citep[][]{fasano87} to quantify the probability that 
the AGN at low relative velocity (the radio galaxy, AGN1110, AGN1317, and AGN1354) and the galaxies without AGN are two different 
populations in the phase-space diagram. 
This statistical test is the generalization of the classical 1D KS test and is suitable for analyzing (random) samples 
defined in two or three dimensions.
Based on the 2D KS test, the significance of the equivalence between the distribution of AGN with low relative velocity and that 
of galaxies without AGN in the phase-space diagram is 23$\%$. 
In other words, the probability is low that the two populations come from the same parent distribution.   
Different regions in the phase-space diagram are populated by galaxies with different infall times, that is, 
they have accreted onto the cluster at different epochs. 
Galaxies tend to follow a path in this diagram as they settle into the cluster potential, 
and those that have spent more time in the structure settle at low relative velocities \citep[see Fig.~1 in][]{Rhee17}. 

The other three AGN of our sample are located in different regions of the phase-space diagram. 
Galaxy AGN1206 is located within $R_{200}$ ($ r/R_{200} \simeq 0.3$) and has a different velocity from the systemic velocity of the protocluster 
($\Delta$v/$\sigma_{\rm v} \simeq -1.4$), while 
$sBzK.7556$ and $sBzK.5860$ are located outside $R_{200}$ ($ r/R_{200} \simeq 1.1$ for both AGN), and their velocity is slightly different with respect to the systematic 
velocity of the protocluster ($\Delta$v/$\sigma_{\rm v} \simeq 0.4$ and $\Delta$v/$\sigma_{\rm v} \simeq -0.7$ for $sBzK.7556$ and $sBzK.5860$, respectively).

\begin{figure*}
\centering
\includegraphics[width=0.48\textwidth]{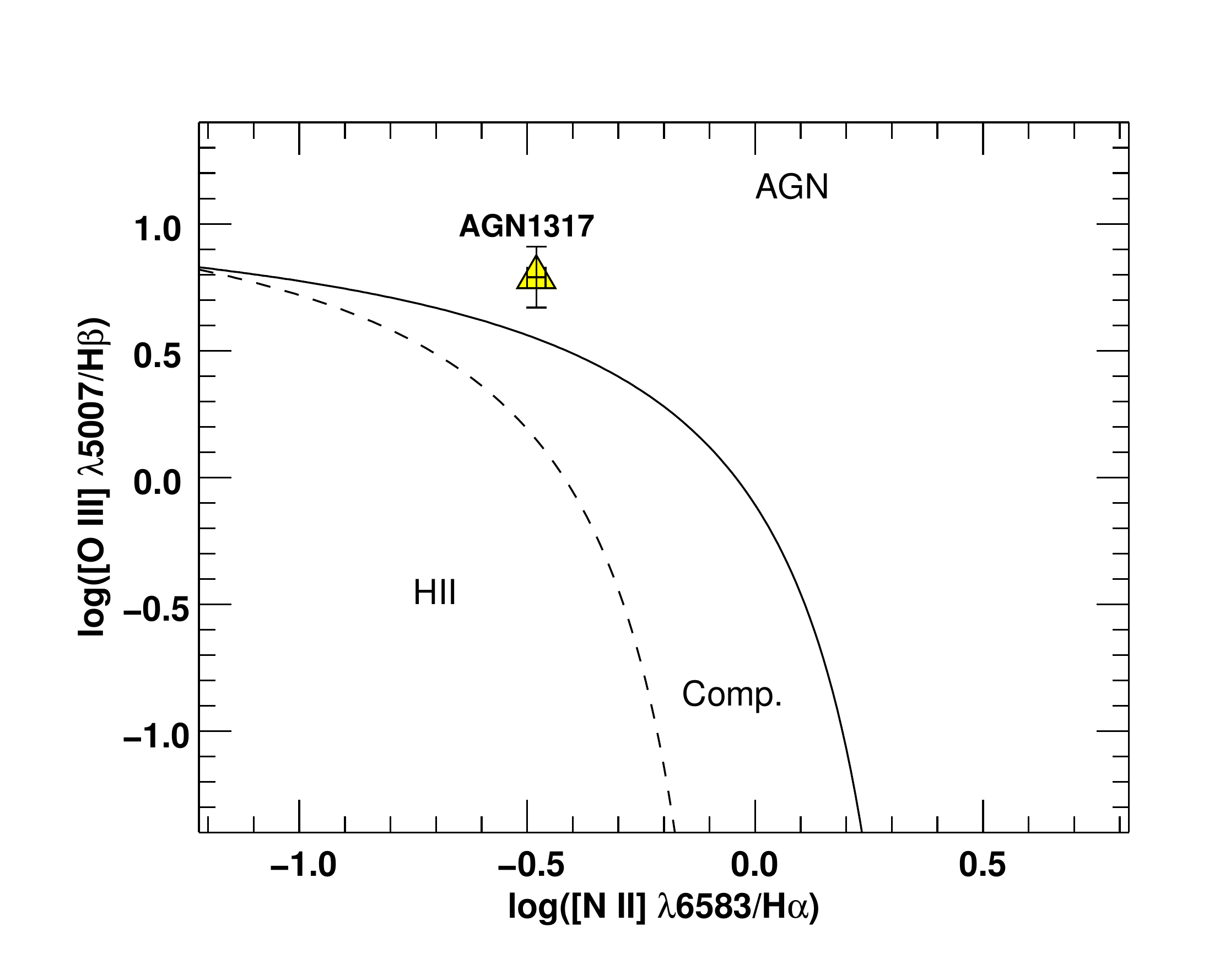}
\includegraphics[width=0.48\textwidth]{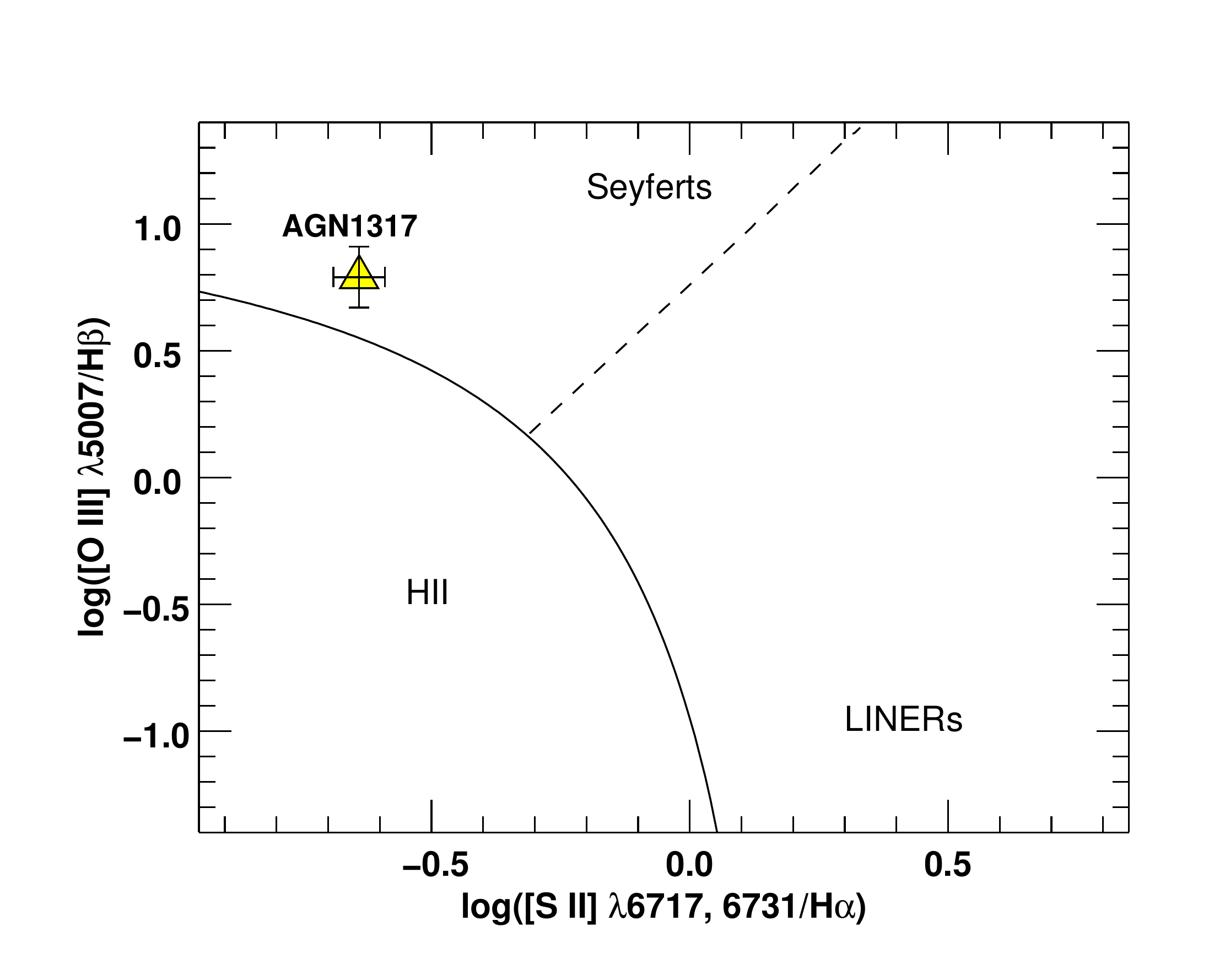}
\caption{
\citet{baldwin81} (BPT) diagrams.
\textit{Left panel:} 
 [N~{\sc ii}]/H$\alpha$ vs. [O~{\sc iii}]/H$\beta$ BPT diagram showing the location of AGN1317 (yellow triangle).
We adopted the classification of the galaxies in  H~{\sc ii}-region-like, composite AGN-H~{\sc ii} types, 
and AGN according to \citet{kewley06}.
The \citet{kewley01} extreme starburst line and the \citet{kauffmann03} pure star formation line are shown as the solid and dashed lines, respectively. 
\textit{Right panel:} 
Same as the left panel for the [S~{\sc ii}]/H$\alpha$ vs. [O~{\sc iii}]/H$\beta$ BPT diagram. 
We adopted the classification of the galaxies in  H~{\sc ii}-region-like, Seyferts, and LINERs according to \citet{kewley06}.
The \citet{kewley01} extreme starburst line and the \citet{kewley06} Seyfert-LINER line are shown as the solid and dashed lines, respectively.   
}
\label{fig:bpt}
\end{figure*}

\section{Characterization of the AGN population}
\label{sec:focus-agn}   
In this section we analyze the characteristics of the AGN in protocluster 7C~1756+6520, 
with a focus on one object in particular, AGN1317.

\subsection{AGN population within the protocluster}
\label{sec:agn}
As described in Sect.~\ref{sec:intro}, the galaxy protocluster associated with 7C~1756+6520 
is characterized by a high fraction of AGN protocluster members:
seven AGN, including the central radio galaxy, have been spectroscopically confirmed in close proximity    
both spatially and in redshift space of the protocluster.
This high AGN fraction detected so far, $\sim$23$\%$, makes the overdensity around 7C~1756+6520 similar 
to the interesting and well-studied cluster around the radio galaxy PKS~$1138-262$ from this point of view as well
\citep[][]{pentericci02,croft05}, in addition to the extension (see Sect.~\ref{sec:gal-distr}).
The source PKS~$1138-262$ is a massive forming radio galaxy at $z \sim 2.16$ \citep[][]{pentericci98} that
is surrounded by overdensities of Ly$\alpha$ emitters \citep[][]{pentericci00}, extremely red objects \citep[][]{kurk04a}, 
H$\alpha$ emitters \citep[][]{kurk04b},  and X-ray emitters \citep[][]{pentericci02}, several of which are 
spectroscopically confirmed to be close to the radio galaxy redshift. 
Five of the 18 X-ray sources ($\sim$28$\%$) detected by \citet{pentericci02} are AGN, including the central radio galaxy. 
From the soft X-ray luminosity function of AGN, \citet{pentericci02} estimated how many sources are expected
in a given region of the cluster PKS~$1138-262,$ finding that it contains about twice the number of expected AGN. 
More recently, \citet{pentericci13} also found high AGN fractions by studying eight galaxy groups from $z \sim 0.5$ to $z \sim 1.1$.   
They found that the fraction of AGN with Log~$L_{H} > 42$~erg~s$^{-1}$ in galaxies with $M_R < -20$ varies from less than $5\%$ to
$22\%$, with an average value of $6.3\%$, which is more than double the fraction for massive cluster at similar high redshifts \citep[e.g.,][]{overzier05}.
\citet{martini13} estimated that the cluster AGN fraction in a sample of 13 clusters of galaxies at $1 < z < 1.5$ is $\sim$3$\%$
for AGN with rest-frame, hard X-ray luminosity greater than $L_{X,\,H} \geq 10^{44}$~~erg~s$^{-1}$.
Based on these findings, the galaxy protocluster around 7C~1756+6520 seems to be a particularly interesting object.

The seven spectroscopically confirmed AGN in protocluster 7C~1756+6520 also suggest a higher AGN fraction
than in local and moderate high-redshift ($z \sim 0.6$) galaxy clusters \citep[e.g.,][]{pentericci02,eastman07,pentericci13}.
In the nearby Universe, AGN rarely occur in clusters, comprising only $\sim$1\% of all cluster galaxies 
\citep[e.g.,][]{dressler85,martini07,arnold09,martini07,martini09}, 
while they are much more common in the field population \citep[$\sim$5\%, e.g., ][]{huchra92,deng12}.
As described in Sect.~\ref{sec:intro},
some works found a clear increasing trend of the AGN fraction with redshift. 
\citet{pentericci13} reported that most of the low-redshift groups have no AGN, while at $z > 0.5-1,$ many have an AGN fraction of $\sim$5--10$\%$ among bright galaxies
\citep[see also][]{eastman07,galametz09b,martini09,martini13}.    

The high fraction of AGN in protocluster 7C~1756+6520 might suggest that the AGN were probably 
triggered at around the same time, presumably by the ongoing formation of the cluster. 
This would support models in which AGN feedback is an important component of the early phases of galaxy 
and cluster formation \citep[e.g.,][]{haehnelt00,rawlings03}.

\subsection{Focus on AGN1317}
\label{sec:agn.1317} 
Among the spectroscopically confirmed AGN proto-cluster members, AGN1317 seems to be particularly interesting 
based on the rich observational dataset collected so far on it (Galametz~et al.~2009, G10, M12, Casasola~et al.~2013, and this work). 
The galaxy AGN1317 has been classified as an AGN for the first time based on the detection 
of broad Mg~{\sc ii}, [Ne~{\sc v}], and [O~{\sc ii}] by G10.
Based on the emission lines we observed, we are able to better characterize the AGN nature of AGN1317 
by using Baldwin, Phillips \& Terlevich (BPT) diagrams \citep{baldwin81}.
These diagrams are typically used to distinguish between normal AGN (Seyferts and quasars), low-ionization nuclear emission-line regions (LINERs), 
and normal H~{\sc ii} regions  on the basis of their [O~{\sc iii}]$\lambda$~5007~\AA/H$\beta$, [N~{\sc ii}]$\lambda$~6583~\AA/H$\alpha$,  
[S~{\sc ii}]$\lambda\lambda$~6716, 6731~\AA/H$\alpha$, and [O~{\sc i}]$\lambda$~6300~\AA/H$\alpha$ flux ratios. 

We produced the BPT diagrams [O~{\sc iii}]/H$\beta$ versus [N~{\sc ii}]/H$\alpha$ and
[O~{\sc iii}]/H$\beta$ versus [S~{\sc ii}]/H$\alpha$ by    
combining observations presented in this paper and in M12.
Since we treated ratios of nearby pairs of lines  ([O~{\sc iii}] and H$\beta$, and [N~{\sc ii}] and H$\alpha$), 
differences in flux calibration that are due to different observation sets (M12 and ours)  did not influence the location of the AGN in the 
diagnostic diagrams.  
Unfortunately, we were not able to separate the H$\beta$ narrow and broad components (but we measured the flux from both components,  
H$\beta$ n+b, see Table~\ref{tab:fluxes}), and for this reason, we estimated the expected flux of the H$\beta$~n component 
assuming that the H$\beta$~b/H$\beta$~n flux ratio is the same as for H$\alpha$~b/H$\alpha$~n ($\sim$3.5).   
This gave an [O~{\sc iii}]/H$\beta$ flux ratio of $\sim$6.2.
From M12 we derived a [N~{\sc ii}]/H$\alpha$ flux ratio of $\sim$3.3.  

Figure \ref{fig:bpt} shows the location of AGN1317 in the two BPT diagrams we produced.
We adopted the galaxy classification of \citet{kewley06}.
In the left panel of Fig.~\ref{fig:bpt} ([O~{\sc iii}]/H$\beta$ vs [N~{\sc ii}]/H$\alpha$), the solid and dashed lines 
are the classification curves of \citet{kewley01} and \citet{kauffmann03}, respectively.
Galaxies that lie below the dashed \citet{kauffmann03} line are classified as H~{\sc ii}-region-like galaxies.
Star-forming galaxies form a tight sequence from low metallicities
(low [N~{\sc ii}]/H$\alpha$, high [O~{\sc iii}]/H$\beta$) to high metallicities (high [N~{\sc ii}]/H$\alpha$, 
low[O~{\sc iii}]/H$\beta$), which we refer to as the star-forming sequence.
The AGN mixing sequence begins at the high-metallicity end of the star-forming sequence and extends toward high 
[O~{\sc iii}]/H$\beta$ and [N~{\sc ii}]/H$\alpha$ values.
Galaxies that lie in between the two classification lines are on the AGN-H~{\sc ii} mixing sequence and 
are classified as composites.
Galaxies that lie above the \citet{kauffmann03} line are classified as AGN.
Based on this classification and our emission-line flux ratios, we confirm that galaxy AGN1317 is an AGN.     
We infer the same result with the BPT diagram shown in the right panel of Fig.~\ref{fig:bpt} ([O~{\sc iii}]/H$\beta$ vs. [S~{\sc ii}]/H$\alpha$).  
As for the left panel, the solid line represents the \citet{kewley01} extreme starburst classification, while the dashed line 
shows the separation between Seyferts and LINERs. 
Seyfert galaxies lie above the \citet{kewley01} classification line and above the Seyfert-LINER line, 
while LINERs lie above the \citet{kewley01} classification line and below the Seyfert-LINER line.   
Based on the adopted classifications and our emission-line flux ratios, we confirm for the first time through BPT diagrams that  
AGN1317 hosts an AGN (Seyfert or quasar).

The protocluster galaxy AGN1354 with the detection of H$\alpha$, H$\beta$, and [N~{\sc ii}]$\lambda$~6583\AA\ lines (see Table~\ref{tab:z}) 
might be placed in the BPT diagram (left panel of Fig.~\ref{fig:bpt}) by estimating an upper limit for the [O~{\sc iii}]$\lambda$~5007\AA\ line.
Unfortunately, H$\alpha$ and [N~{\sc ii}]$\lambda$~6583\AA\ lines are superimposed to the sky-emission in the spectrum of AGN1354
(see Table~\ref{tab:fluxes}).

\section{Comparison with theoretical predictions}
\label{sec:theory}
The comparison between the observations and the predictions of semi-analytic models (SAMs) that include AGN growth 
can help us to understand the main physical processes that drive the formation and fuel BHs.
The AGN frequency is expected to depend quite strongly on the environment, with factors such as the local galaxy density 
and one-on-one interactions 
\citep[e.g.,][]{pentericci13,sabater13}.

There are two main modes of AGN growth and feedback
in these models: the quasars mode, and the radio mode .
The quasar mode applies to BH growth during gas-rich mergers where the central BH of the main progenitor
grows both by absorbing the central BH of the minor progenitor and by accreting the cold gas. 
This mode corresponds to high-luminosity AGN, which accrete at near their Eddington ratio,
and which therefore are subject to a wind. 
In the radio mode, quiescent hot gas is accreted onto the central super-massive BH, and this accretion comes 
from the surrounding hot halo.
The radio mode corresponds to low-luminosity AGN, typically well below the Eddington rate, that 
emit a radio jet.
This model captures the mean behavior of the BH on timescales much longer than the duty cycle.

We considered two different SAMs: 
the model of  \citeauthor{menci04}~(2004, hereafter M04), and the model implemented in the Millennium Simulations 
(MS) as in \citet{guo11}.
In model M04, which is based on the quasar mode, the accretion of gas onto the central BH is triggered by galaxy encounters 
that do not necessarily lead to bound mergers in common host structures such as clusters and especially groups.    
These events destabilize part of the galactic cold gas and thus feed the central BH, following
the physical modeling developed by \citet{cavaliere00}.    
This model finds that at high redshift, while the protogalaxies grow rapidly by hierarchical merging, fresh gas is imported,
and the BHs are fueled at their full Eddington rates. 
At lower redshift, the dominant dynamical events are galaxy encounters in hierarchically growing groups. 
At this point, refueling diminishes as the residual gas is exhausted, and the destabilizing
also decreases. 

In the MS model for BH growth and AGN feedback, \citet{guo11} followed \citet{croton06},  who implemented both quasar mode 
and radio mode.  
In the quasar mode, BH accretion is allowed during both major and minor mergers, but the efficiency 
in the latter is lower because the mass accreted during a merger depends on various factors, including the ratio between the mass of a member and that 
of the central cluster galaxy.
In the radio mode, the growth of the SMBH is the result of continuous hot gas accretion once a static hot halo has formed around
the host galaxy of the BH. 
This accretion is assumed to be continual and quiescent \citep[see][for details]{croton06}.

The high AGN fraction characterizing protocluster 7C~1756+6520 seems to be more consistent with the predictions of the MS model.
This model predicts a steep increase in the AGN fraction with redshift, both in groups and in clusters \citep[see, e.g., Fig.~4 in][]{pentericci13}, 
which is linked to the marked rise of major mergers, which are the only mergers considered for the quasar mode, toward high redshifts.    
Conversely, the M04 model predicts a milder increase in AGN fraction with redshift because in this model, minor mergers and close encounters 
are also very important, and their frequency does not depend very strongly on redshift, since small dark matter halos continue to merge frequently 
until low redshift.  
The better agreement with the MS model would validate the implemented mode of AGN growth in the model, and in particular would stress the importance
of galaxy encounters, not necessarily leading to mergers, as an efficient AGN triggering mechanism.
However, we recall that the model M04 successfully reproduces the observed properties of both galaxies and AGN across a wide
redshift range \citep[$z \sim 0-4$, e.g., ][]{fontana06,menci08,calura09,lamastra10}.

Alongside galaxy mergers, the ram pressure that cluster galaxies experience as they move through the intracluster medium
may induce AGN activity in addition to star formation \citep[e.g.,][]{poggianti17,marshall18,ramos-martinez18}. 
This pressure is able to strip gas from galaxies (ram pressure stripping), which may cause tails of stripped gas to form behind 
the galaxy as it moves in the cluster, as has been observed for galaxies in the Virgo cluster 
\citep[e.g.,][]{kenney04,crowl05,chung07}.
Ram pressure stripping leads to a decreased prevalence of radio-mode AGN activity in the centers of clusters
\citep[e.g.,][]{ellison11,ehlert14,khabiboulline14}, since the ram pressure has depleted the gas supply 
of these central galaxies.
However, models and hydrodynamical simulations show that lower ram pressures can compress the gas in the galaxy, and
may induce star formation \citep[e.g.,][]{fujita99,kronberger08,kapferer09,tonnesen09,bekki14}, which is also supported by observations 
\citep[e.g.,][]{lee17}.
These moderate ram pressures might conceivably also lead to higher BH accretion, and hence trigger AGN activity,
since ram pressure can lead to angular momentum loss in gas clouds \citep[][]{tonnesen09} and trigger gravitational instability in the
galactic disk \citep[][]{schulz01}. This in turn may lead to gas being deposited into the galaxy center.
Ram-pressure- and merger-induced AGN activities are both dependent on the galaxy location within the cluster.

Recently, \citet{marshall18} has performed hydrodynamical simulations of the effect of ram pressure on gas-rich galaxies, finding that 
below the regime of ram pressure stripping, the enhanced pressure might lead to an elevated level of star formation and the onset of AGN activity.  
They have tested this effect with an SAM based on the MS and compared it to an observational sample of galaxies
in low- and high-redshift clusters.
The phase-space properties of observed AGN populations are consistent with a triggering scenario for intermediate ram pressures ($P_{\rm ram}$),
$2.5 \times 10^{-14} < P_{\rm ram} < 2.5 \times 10^{-13}$. 
The critical range corresponds to ram pressures expected around the virial radius in low-redshift galaxy
clusters and at smaller radii in high-redshift ones \citep[see Fig.~6 in][]{marshall18}. 
The observational AGN fractions show a slower decrease with radius than the $z = 1$ model of \citet{marshall18}, 
while the location of the peak is consistent with the model \citep[see Fig.~9 in][]{marshall18}.
The observations also show a second peak in the distribution of AGN fractions with radius, which is absent in the
$z = 1$ model of \citet{marshall18}.
For protocluster 7C~1756+6520, this model would mean that 4 of the 7 AGN are located within $\sim$$0.5 \times r_{\rm {vir}}$ (see Fig.~\ref{fig:space-phase}), 
which appears to be in agreement with the first peak of the radial distribution of the AGN fractions predicted by 
\citet{marshall18} and observed by \citet{martini09}. 
The definition of an outer AGN peak and of a trend at larger radii is prevented by the small number of protocluster members identified so far (31).
Following \citet{martini09}, we conclude that the basic comparison of our sample and other samples of AGN 
belonging to high-redshift clusters
shows that the ram pressure triggering model is at least broadly consistent with the observations.

\section{Summary and conclusions}
\label{sec:conclusions} 
We presented new near-IR LBT observations of galaxies belonging to the field of the protocluster that is
associated with the radio galaxy 7C~1756+6520 at $z = 1.4156$.      
We observed the central part of the protocluster, including the radio galaxy, three spectroscopically confirmed  
AGN, and other objects that are possible protocluster members.
We also presented an analysis dedicated to the AGN population identified within the protocluster.
Our main conclusions are listed below.
\begin{itemize}
\item Most of the redshift identification is based on the detection of the H$\alpha$, H$\beta$, 
and [O~{\sc iii}]$\lambda$~5007~\AA\ emission lines.  
For four galaxies that were previously targeted as protocluster members, we derived the redshift by detecting emission lines that have never been detected before.
We identified a new protcluster member, by detecting the H$\alpha$ and [O~{\sc iii}]$\lambda$~5007~\AA\ emission lines,
and eight new possible protocluster members by assuming that the only detected emission line 
is the [O~{\sc iii}]$\lambda$~5007~\AA\ or the H$\beta$ line.
If the emission-line assignation is correct, these galaxies should belong to the redshift range of 
$z \sim 1.38 - 1.43$.
The stacked spectrum of the possible protocluster members in which we detected the [O {\sc iii}]$\lambda$~5007~\AA emission line
revealed the second line of the [O {\sc iii}] doublet at 4959~\AA\ and of H$\beta$, 
which confirms that they belong to the protocluster. 

\item
By merging the spectroscopically confirmed protocluster members derived from this work, G10, and M12, 31 galaxies, 
including the central radio galaxy, are found around redshift $1.4152 \pm 0.056$. This corresponds to 
peculiar velocities $\lesssim$5000~km~s$^{-1}$ with respect to the radio galaxy.
The velocity dispersion of the sources within 2~Mpc from the radio galaxy 
is $\sigma_{\rm {cl}} \sim 1700 \pm 120$~km~s$^{-1}$.
\item 
Based on the identified 31 proto-cluster members, the redshift distribution could be described by a Gaussian function 
with respect to the main peak at $z \sim 1.42$. 
The application of the robust statistical DS test shows that we are not able to define the presence of substructures 
within the protocluster.

\item
The phase-space diagram shows that three AGN are located in the region  
at low velocity and small radii, which together with the central radio galaxy corresponds to galaxies with a high 
probability to be a  virialized population that has been 
part of the overdensity for a long time. 

\item
A peculiar property of this protocluster is its AGN fraction, which at $23\%$ is higher than what typically characterizes
low-, moderate-, and high-redshift clusters.    
The high AGN fraction and the distribution of the AGN within the protocluster is broadly consistent with predictions 
of some theoretical models on the AGN feedback, based on galaxy interactions and ram pressure. 

\item
By combining our observations with other available observations, we confirmed for the first time through the BPT diagrams
that AGN1317 hosts an AGN.
\end{itemize}

We plan to analyze high-resolution (1.5 and 5 mas) radio data (at 6 and 8~cm) of the central H$z$RG 
that we have already acquired with the EVN (European VLBI\footnote{VLBI is the acronym of Very Long Baseline Interferometry.} Network) 
and $e$-MERLIN/VLBI facilities. 
These observations have been motivated by the evidence that H$z$RGs in the center of clusters are the probable ancestors 
of the brightest cluster galaxies (BCGs). BCGs are objects that are typically located at the center of clusters and are considered the most luminous, massive, 
and extended sources in the Universe.
Because of the dominant role inside clusters and the peculiar multiwavelength properties of the BCGs, the evolution of these objects 
through cosmic time is particularly interesting. 
These centimeter observations will allow us to establish the radio structure and properties of the H$z$RG, identify its core and resolve 
the components that have been detected with the Very Large Array. It will also allow us to map the subarcsecond-scale emission with a simultaneous 
multi-frequency space-resolved spectral analysis and connect the radio emission at all angular scales.
We furthermore plan to propose new NIR observations to complete the data presented here in order to clearly determine the spectroscopic redshift of the 
possible cluster members found so far and to search for new members.

\begin{acknowledgements}
We are grateful to the anonymous referee, whose comments and suggestions greatly improved the quality of this manuscript.
V.~C. acknowledges the DustPedia financial support.
DustPedia is a collaborative focused research project supported by the European Union under the
Seventh Framework Programme (2007--2013) call (proposal No. 606824). 
The participating institutions are: Cardiff University, UK; National Observatory of Athens, Greece; 
Ghent University, Belgium; Universit{\'e} Paris Sud, France; 
National Institute for Astrophysics, Italy and CEA (Paris), France.
J.~F. acknowledges the financial support from UNAM-DGAPA-PAPIIT IA104015 grant, Mexico.
This research has made use of the SIMBAD database, operated at CDS, Strasbourg, France \citep[][]{wenger00}.
This research has made use of the NASA/IPAC Extragalactic Database (NED) which is operated by the Jet Propulsion Laboratory, 
California Institute of Technology, under contract with the National Aeronautics and Space Administration. 
The authors thank A.~Galametz for making available the $B$-band image of the field around the radio galaxy 7C~1756+6520,
which has previously been used in \citet{vivi13}.
\end{acknowledgements}


\end{document}